\documentclass[psfig,epsf]{article}
\usepackage{multicol}
\usepackage{epsfig}
\begin{document}

\def\Journal#1#2#3#4{{#1} {\bf #2} (#3) #4}
\def \PRL      {Phys. Rev. Lett.~}
\def \PR       {Phys. Rev.}
\def \PRD      {Phys. Rev. D}
\def \PL       {Phys. Lett.~}
\def \PLB      {Phys. Lett. B}
\def \ZPC      {Z. Phys. C}     
\def \NPB      {Nucl. Phys. B}
\def \NPD      {Nucl. Phys. D}
\def \PR       {Phys. Rep.~}
\def \INC      {Il Nuovo Cimento}
\def \NIM      {Nucl. Instrum. Methods}
\def \NIMA     {Nucl. Instrum. Methods Phys. Res. Sect. A}
\def \CPC      {Comput. Phys. Commun.}
\def \EUR      {Eur. Phys. J. C}
\def \PTP      {Prog. Theor. Phys.}
\def \PTPS     {Prog. Theor. Phys. Suppl.}
\def \JHEP     {J. High Energy Phys.}
\def \etal     {\relax\ifmmode{et \; al.}\else{$et \; al.$}\fi}
\def \ANN      {Ann. Rev. Nucl. Part. Sci.}
\def \JAP      {J. Phys. Soc. Jap.}
\def \SOVJ     {Sov. J. Nucl. Phys.}

\begin{center}
{\Large \bf Recent Results of High $p_{T}$ Physics at the CDF II} \\

\vspace{4mm}

Soushi Tsuno (CDF Collaboration) \\
Department of Physics, Faculty of Science, Okayama University \\ 
3-1-1 Tsushima-naka, Okayama 700-8530, Japan \\
\end{center}

\begin{abstract}
The Tevatron Run II program is in progress since 2001. The CDF experiment 
have accumulated roughly five times more data than in Run I, with much 
improved detectors. Preliminary results from the CDF experiment are presented.
We focus on the recent high $p_{T}$ physics results in Tevatron Run II 
program.
\end{abstract}

\section{Introduction}

The CDF (Collier Detector at Fermilab) experiment is a general purpose 
experiment for the studies of $p\bar{p}$ collisions at the Tevatron Collider 
located at the Fermi National Accelerator Laboratory (Fermilab), in Batavia, 
Illinois, U.S.. The Tevatron accelerator is the highest-energy 
proton-antiproton accelerator machine in the world. The Tevatron accelerator 
\cite{runiihandbook} and the CDF(D0) \cite{tdrrunii} detectors have upgraded 
since the termination of Run I experiment in 1996.

The accelerator complex had added the Main Injector, replacing the old Main 
Ring, to inject higher intensity beams to the Tevatron and to produce more 
anti-protons to be used for collision. The Tevatron beam energy has been 
increased and it resulted in a center-of-mass energy of 1.96 TeV in Run II 
from 1.8 TeV in Run I. The instantaneous luminosity has improved steadily 
since the beginning of Run II, and is marking with a new record. The record 
value at this conference was 6.3 $\times$ 10$^{31}$ cm$^{-2}$s$^{-1}$ which 
was about 3 times higher than the Run-I record value and is close to the 
Run-IIa goal of 8.6 $\times$ 10$^{31}$. The integrated luminosity delivered to 
each experiment has exceeded 450 pb$^{-1}$, and with about 80\% of them 
recorded by the detector. The CDF detector has undergone the upgrade of a 
completely new tracking and forward calorimetry systems, a extension of muon 
tracking, and lots of minor changes as well as DAQ upgrades.

Since the Tevatron Run-II program has officially started in March 2001, the 
CDF is extensively exploring a new physics beyond Standard Model. Measuring 
the high $p_{T}$ phenomena to establish the SM in electroweak sector is also 
important in that it may give us a knowledge in nature about symmetry 
breaking. In this paper, we describe our recent results for the high $p_{T}$ 
physics.

\section{Electroweak Physics}

\subsection{Production of single gauge bosons}

Since the great success of the Standard Model in the recent decades, there has 
been no doubt that the gauge theories are capable to describe the interactions 
between elementary particles. With larger colliding energy available to probe 
higher energy scattering events, a high precision measurement may open a 
beautiful structure among them. 

\begin{figure}[hptb]
\begin{multicols}{2}
\includegraphics[width=8.0cm]{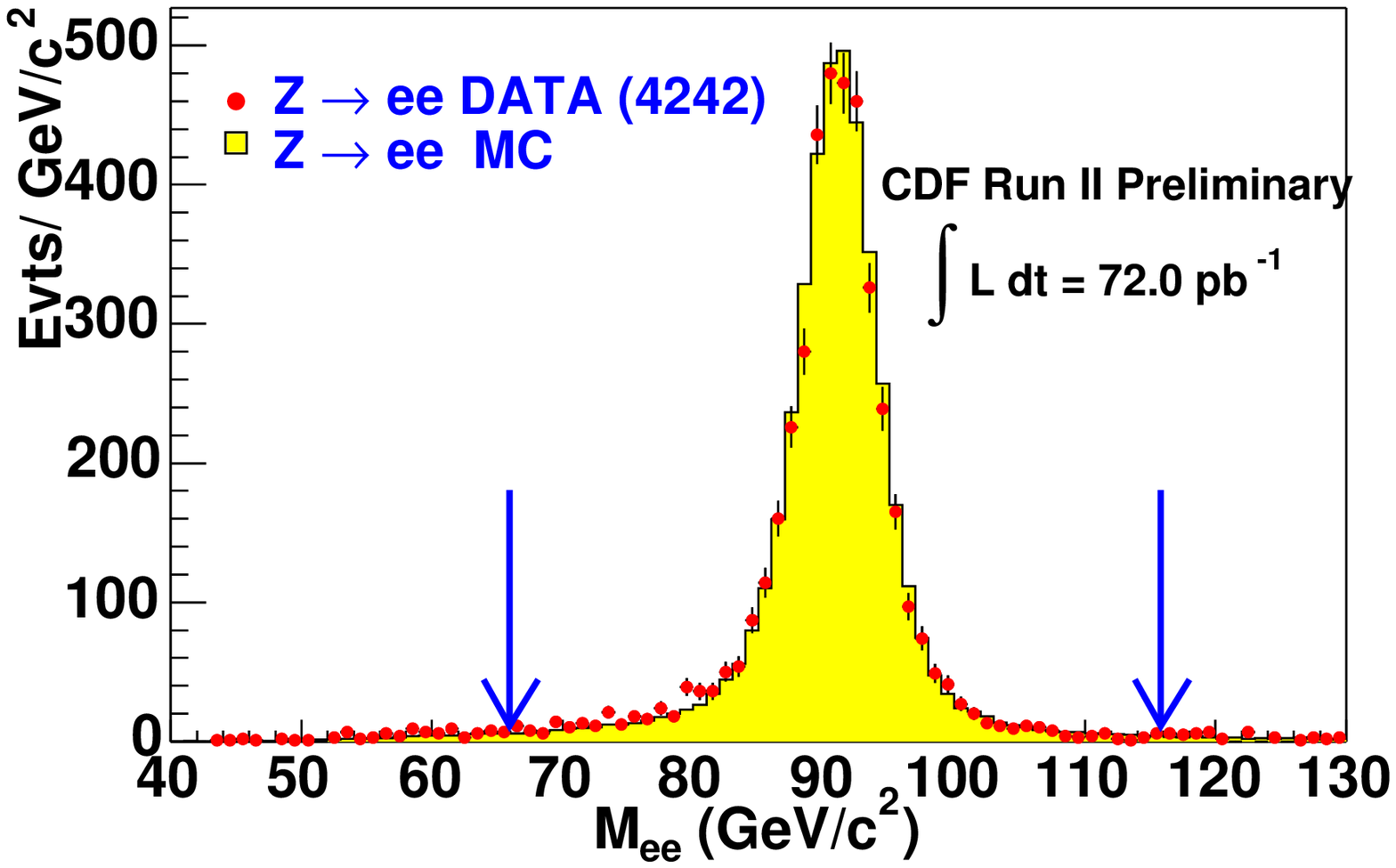}
\caption{Invariant mass distribution of lepton pairs for $Z$ $\rightarrow$ 
$e^{+}e^{-}$ candidates.}
\label{zinvms}
\hspace*{1cm}
\includegraphics[width=5.0cm]{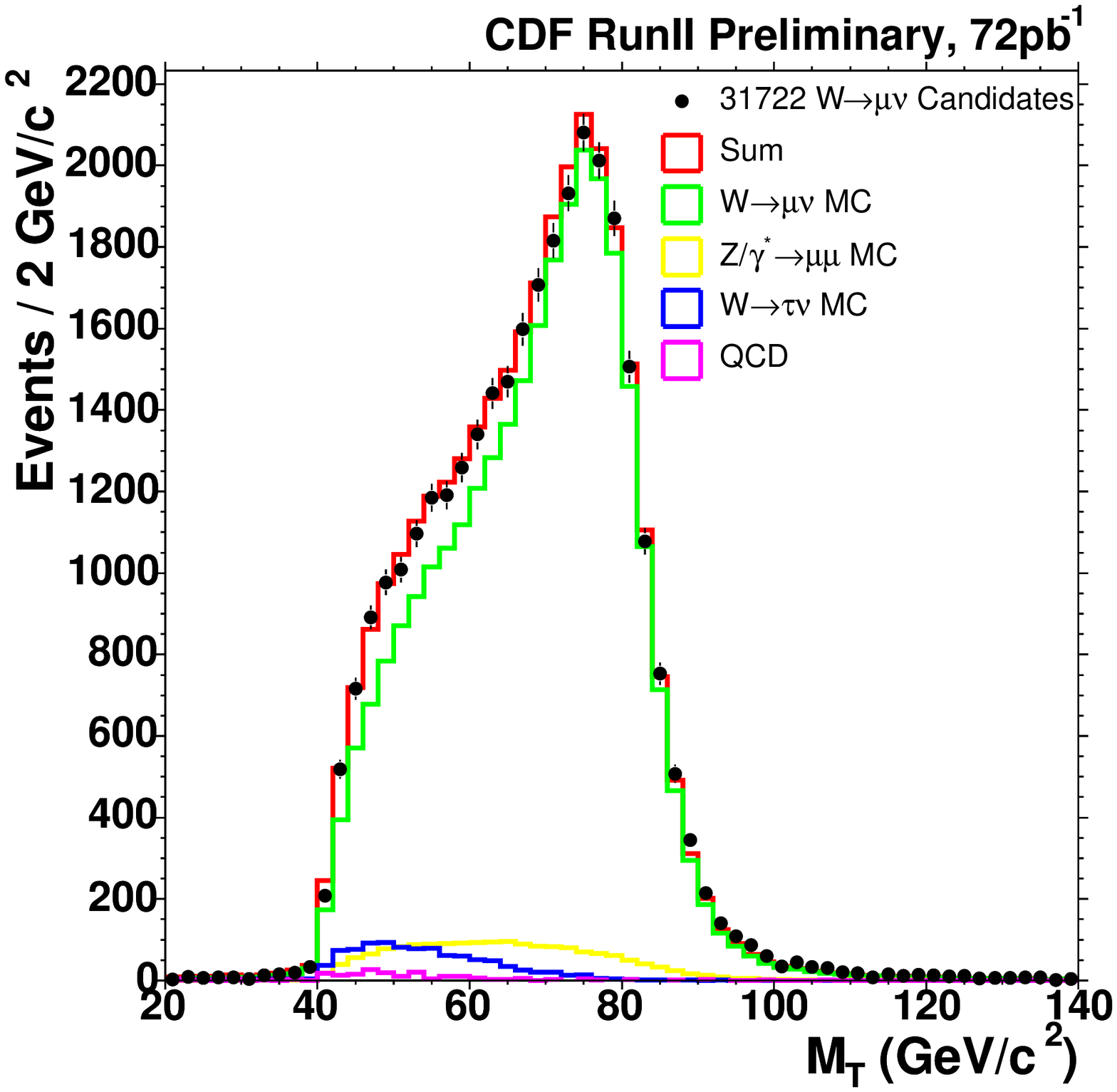}
\caption{Transverse mass distribution of lepton and missing $E_{T}$ system for 
$W$ $\rightarrow$ $e\nu$ candidates.}
\label{wtransms}
\end{multicols}
\end{figure}

The CDF has performed studies of various aspects of the weak boson properties.
They are cleanly identified with their decays to leptons (mostly electrons or 
muons). The leptonicaly decaying Z boson is identified by using two clean 
high $p_{T}$ lepton candidates within the mass range of Z boson. The invariant 
mass distribution for Z boson candidates is shown in Figure \ref{zinvms}. The 
production cross sections are measured to be \cite{0406078}
\begin{eqnarray}
\sigma(p\bar{p} \rightarrow Z/\gamma^{*} \rightarrow \l^{+}\l^{-} + X) = 
254.3 \pm 3.3 (stat.) \pm 4.3 (syst.) \pm 15.3 (lumi.) \quad \mathrm{pb} ,
\nonumber
\end{eqnarray}
in good agreement with a theory prediction of 251.3 $\pm$ 5.0 pb at NNLO level 
with MRST \cite{nnlo}.

Decays of W bosons are distinguished by one high $p_{T}$ lepton and missing 
transverse energy from undetectable neutrino. The transverse mass distribution 
for W boson candidates is shown in Figure \ref{wtransms}, where the transverse 
mass is given as $M_{T}$ $=$ 
$\sqrt{2E_{T}^{\l}E_{T}^{missing}[1-\cos(\Delta\phi^{\l-missing})]}$. The 
production cross sections are measured to be \cite{0406078}
\begin{eqnarray}
\sigma(p\bar{p} \rightarrow W \rightarrow \l\nu + X) = 
2777 \pm 10 (stat.) \pm 52 (syst.) \pm 167 (lumi.) \quad \mathrm{pb} ,
\nonumber
\end{eqnarray}
in good agreement with a theory prediction of 2687 $\pm$ 54 pb at NNLO level 
with MRST \cite{nnlo}.

\begin{figure}[hptb]
\begin{multicols}{2}
\includegraphics[width=7.0cm]{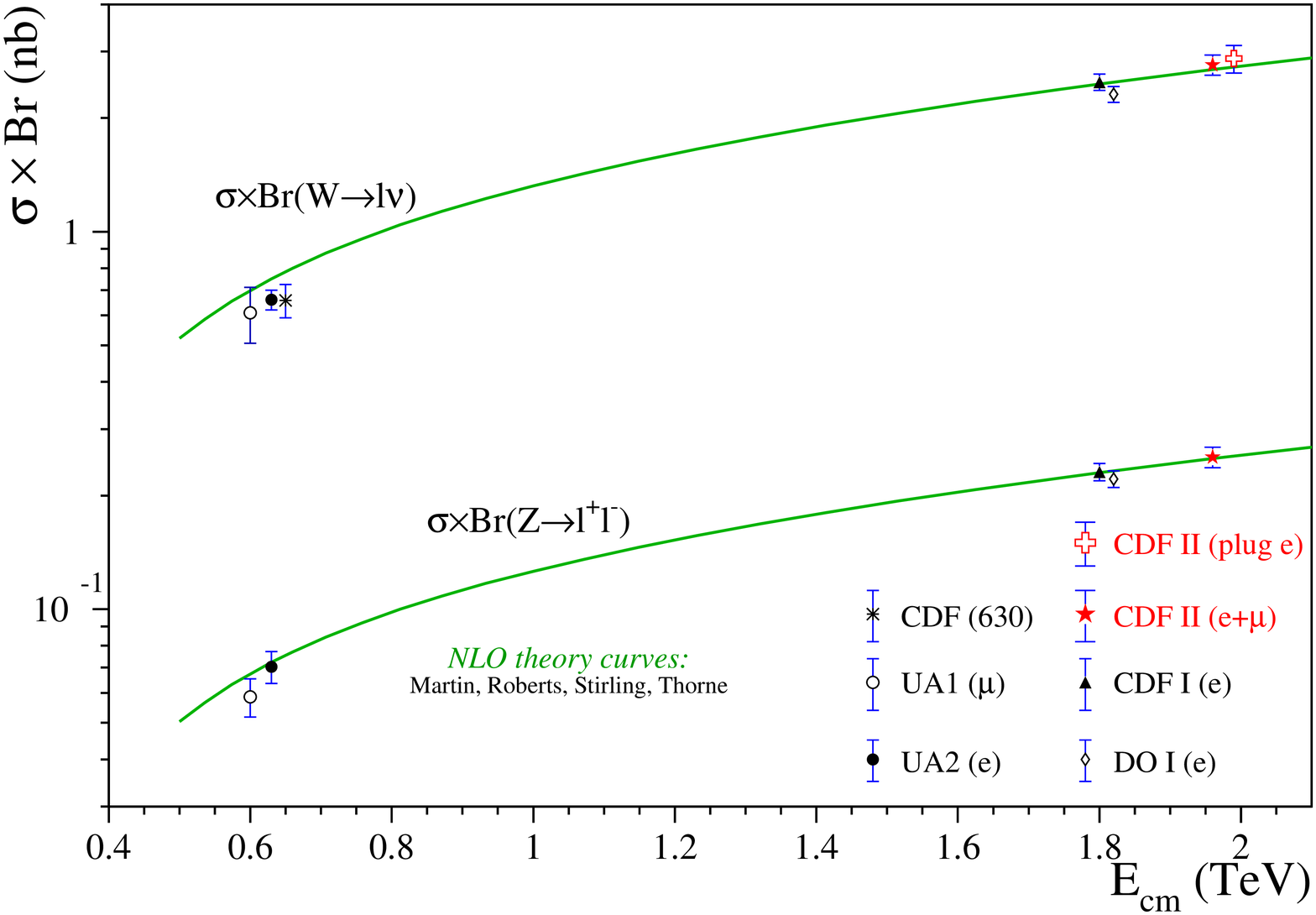}
\caption{Production cross sections of Z and W bosons as a function of 
collision center-of-mass energy.}
\label{cs_vs}
\hspace*{2cm}
\includegraphics[width=4.0cm]{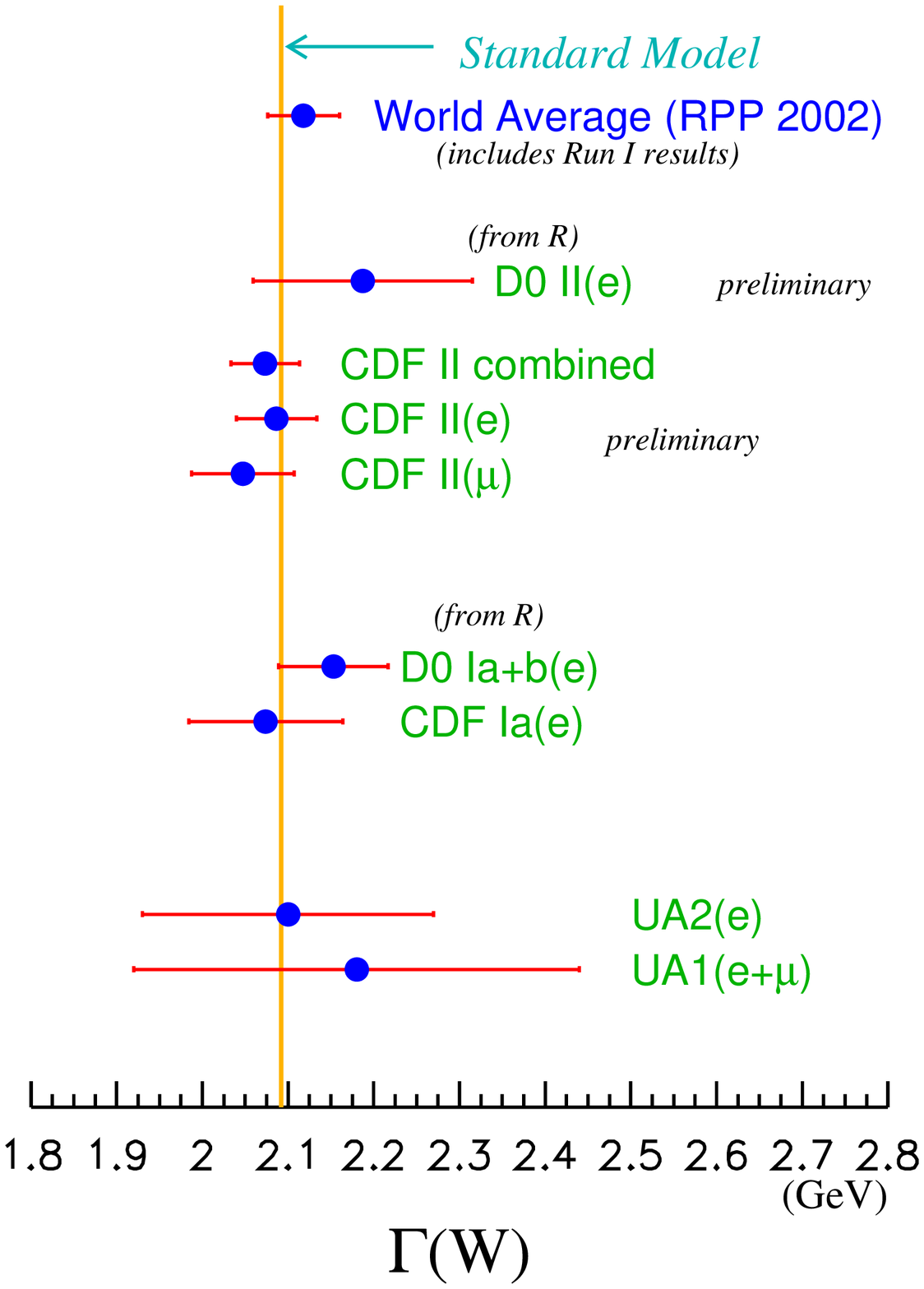}
\caption{W boson width measurements.}
\label{gamma_s}
\end{multicols}
\end{figure}

In Figure \ref{cs_vs}, we summarize the single boson production cross sections
as a function of the center-of-mass energy starting from earlier measurement
at CERN together with our previous measurements of Run I. We can see a good 
agreement with the theory predictions.

The ratio of the Z and W boson production cross sections is used as an 
indirect measurement of the W boson mass and width. The ratio is defined as 
\begin{eqnarray}
R \equiv \frac{\sigma(p\bar{p} \rightarrow W \rightarrow \l\nu + X)}
{\sigma(p\bar{p} \rightarrow Z/\gamma^{*} \rightarrow \l^{+}\l^{-} + X)} = 
10.93 \pm 0.15 (stat.) \pm 0.13 (syst.) .
\nonumber
\end{eqnarray}
Using a theoretical calculation of the ratio of production cross sections and
a measured value of the branching ratio of Z boson at LEP 
experiment \cite{lepz}, we can extract the leptonic branching ratio or total 
width of W boson,
\begin{eqnarray}
Br.(W \rightarrow \l\nu) = 0.1089 \pm 0.0022 \quad , \quad 
\Gamma_{W} = 2.071 \pm 0.040 \quad \mathrm{GeV} .
\nonumber
\end{eqnarray}
The current world average for the W boson width is shown in 
Figure \ref{gamma_s}.

One of remarkable achievements in CDF II experiment is that the track 
reconstructions are possible at the trigger level. As a result of this, tau 
candidate events are able to be triggered, then statistically accessible to 
study various tau lepton physics. Reconstruction of the tau decay is 
accomplished by finding an energy cluster in the calorimeter with isolated 
tracks matched to it. Individual $\pi^{0}$ particles from the tau decay are 
reconstructed using detectors in the calorimeter at shower max, and the 
combined invariant mass of the cluster is required to be less than the mass of 
the tau. Obtained production cross section of Z and W decaying into tau 
leptons are \cite{elwkresults}
\begin{eqnarray}
\sigma(p\bar{p} \rightarrow Z/\gamma^{*} \rightarrow \tau^{+}\tau^{-} + X) = 
242 \pm 48 (stat.) \pm 26 (syst.) \pm 15 (lumi.) \quad \mathrm{pb} ,
\nonumber \\
\sigma(p\bar{p} \rightarrow W \rightarrow \tau\nu + X) = 
2.62 \pm 0.07 (stat.) \pm 0.21 (syst.) \pm 0.16 (lumi.) \quad \mathrm{nb} .
\nonumber
\end{eqnarray}

The lepton universality can be measured with three generations of lepton 
family. Calculating the R ratio for each channel separately, we can derive the 
ratio of the coupling constants,
\begin{eqnarray}
\frac{g_{\mu}}{g_{e}} = 1.011 \pm 0.018 \quad , \quad 
\frac{g_{\tau}}{g_{e}} = 0.99 \pm 0.04 .
\nonumber
\end{eqnarray}
These are all consistent with the SM predictions.

\begin{figure}[hptb]
\begin{multicols}{2}
\includegraphics[width=7.0cm]{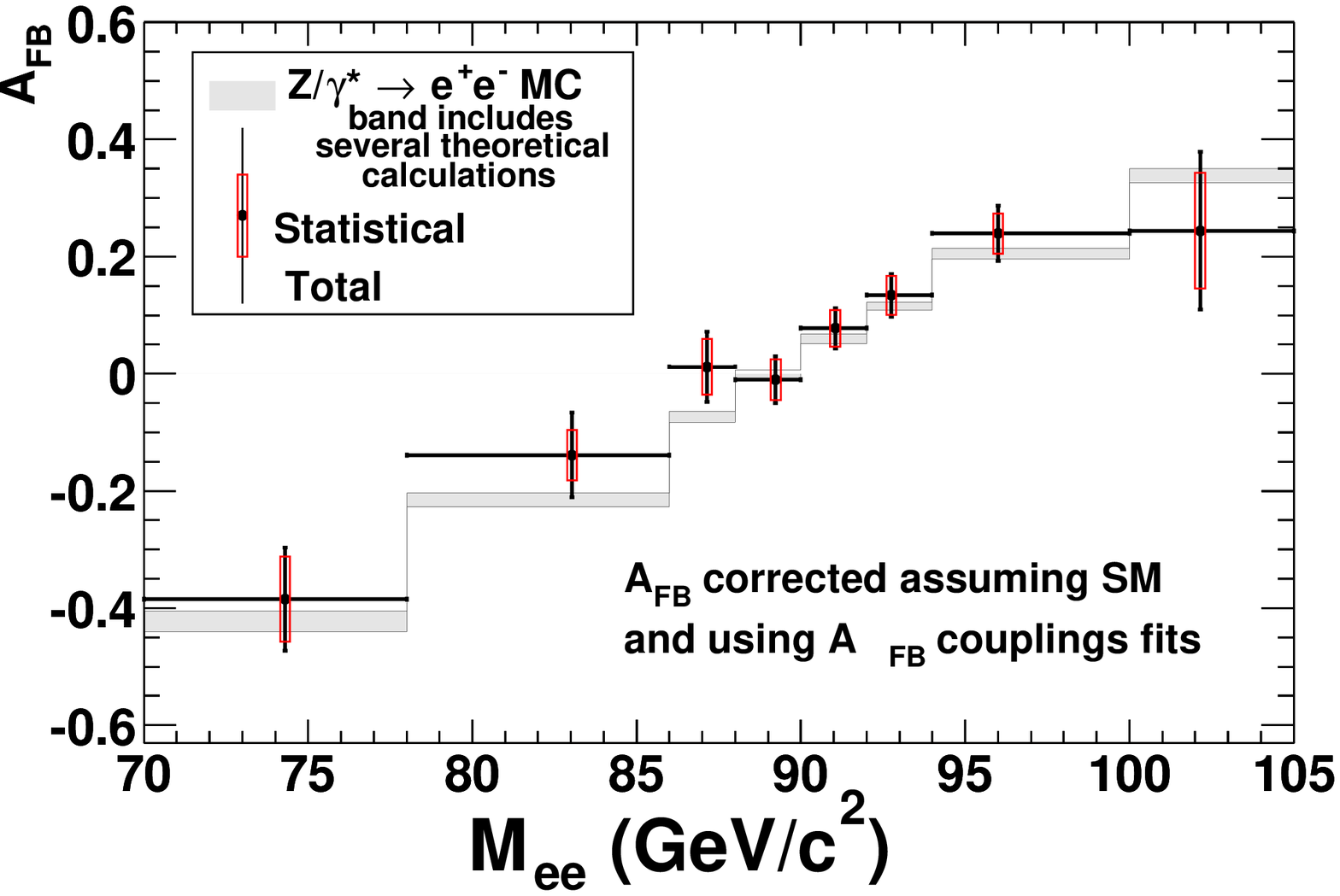}
\caption{Asymmetry distribution of $Z$ $\rightarrow$ $e^{+}e^{-}$ events
around Z-pole.}
\label{afb_zoom}
\includegraphics[width=7.0cm]{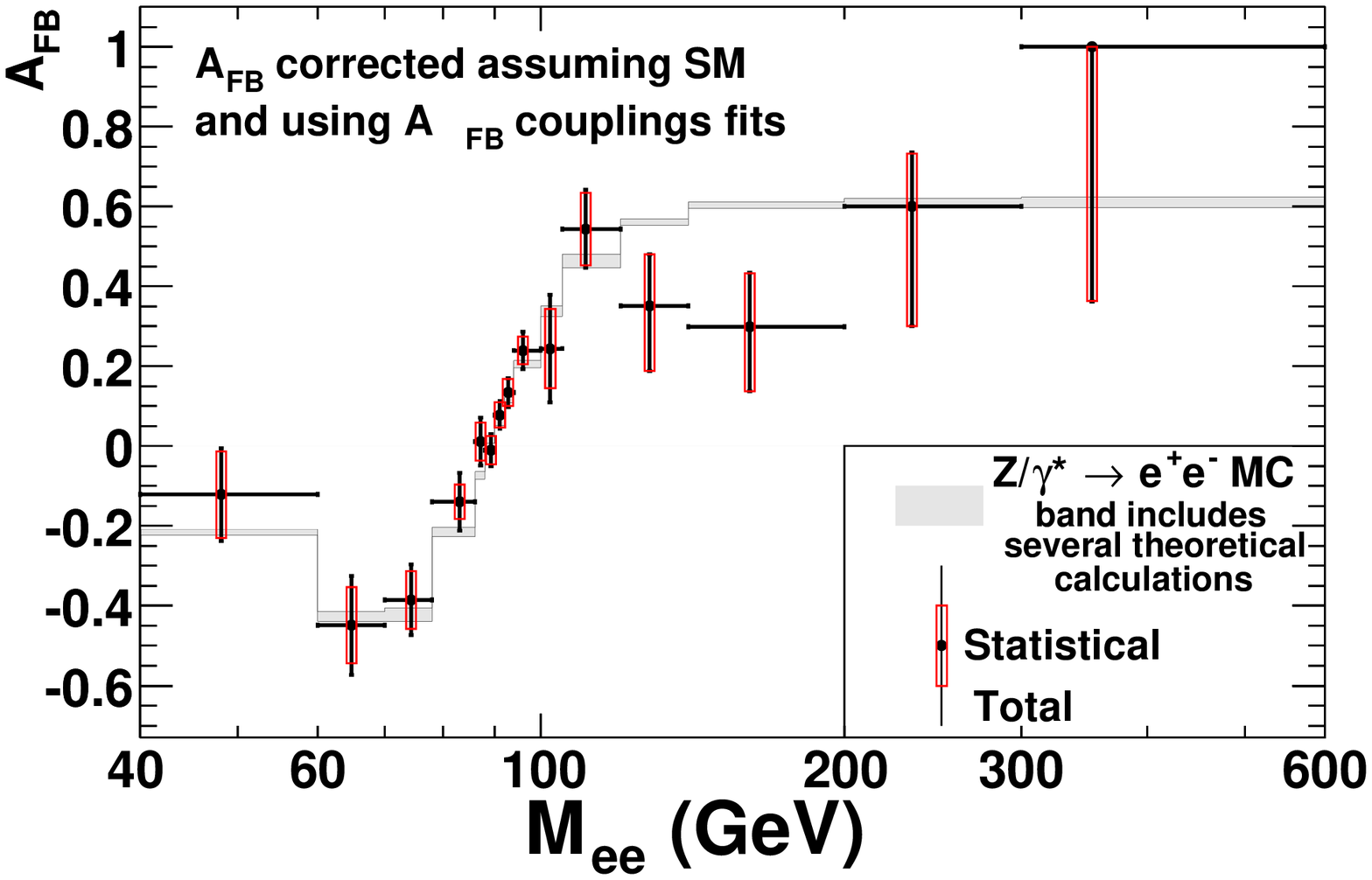}
\caption{Asymmetry distribution of $Z$ $\rightarrow$ $e^{+}e^{-}$ events
in high mass region.}
\label{afb}
\end{multicols}
\end{figure}

In Z boson production, the forward backward asymmetry yields a measurement of 
$\sin^{2}\theta_{W}$ (mixture with up- and down-type quark in PDF) and a 
search for higher mass Z' bosons. This is unique tests only at the Tevatron. 
The asymmetry is defined as 
\begin{eqnarray}
A_{fb} \equiv \frac{\sigma(\cos\theta > 0) - \sigma(\cos\theta < 0)}
{\sigma(\cos\theta > 0) + \sigma(\cos\theta < 0)} ,
\nonumber
\end{eqnarray}
where $\theta$ is the polar angle between the incoming proton and the outgoing 
lepton. The distributions of the asymmetry around Z-pole and in high mass 
region are shown in Figure \ref{afb_zoom} and \ref{afb}, respectively. Both 
distributions agree with the SM predictions. Extracting $\sin^{2}\theta_{W}$ 
from those shape, we can obtain \cite{zasym}
\begin{eqnarray}
\sin^{2}\theta_{W} = 0.2238 \pm 0.0046 (stat.) \pm 0.0020 (syst.) .
\nonumber
\end{eqnarray}

\subsection{Pair production of gauge bosons}

\begin{figure}[hptb]
\begin{multicols}{2}
\includegraphics[width=5.5cm]{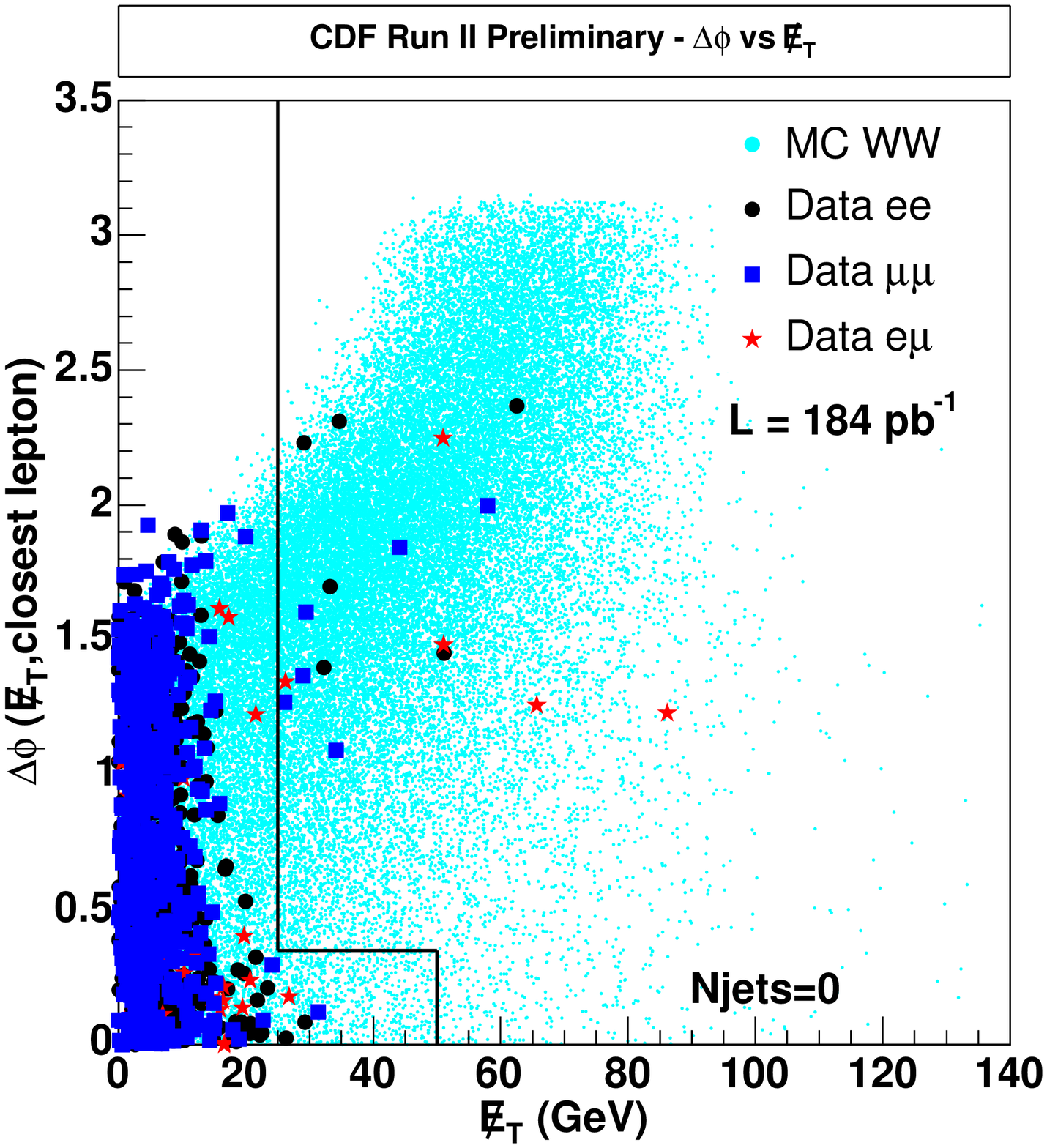}
\caption{Azimuthal angle between the missing transverse energy and the closest 
lepton for it.}
\label{wpairdphimet}
\includegraphics[width=6.0cm]{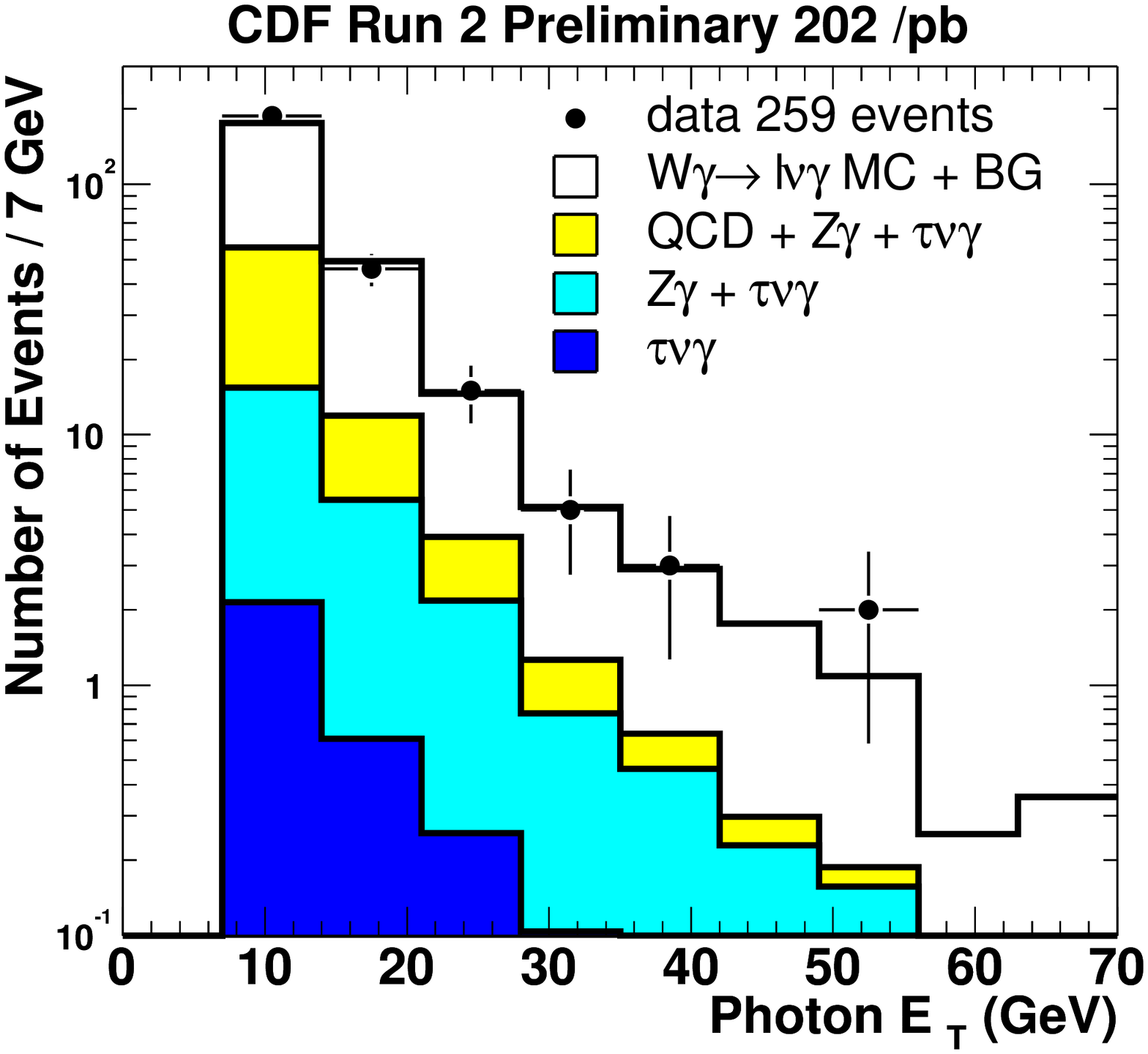}
\caption{Photon transverse momentum for W$\gamma$ production.}
\label{wgammaphoet}
\end{multicols}
\end{figure}

An non-abelian gauge theory describing the electroweak interactions induces 
the three- and four-point self-couplings of gauge bosons. Measurements of the 
pair production of the gauge bosons is a fundamental tests for those 
self-couplings. The cross section for $W^{+}W^{-}$ productions has been 
measured in the {\it dilepton} channel at CDF \cite{wpair}. The azimuthal 
angle between the missing transverse energy and the closest lepton for it is 
uesd to enhance the candidate events, which is shown in 
Figure \ref{wpairdphimet}. The measured value is
\begin{eqnarray}
\sigma(p\bar{p} \rightarrow W^{+}W^{-} \rightarrow \l^{+}\l^{-}\nu\bar{\nu} + 
X) = 14.3 ^{+5.6}_{-4.9} (stat.) \pm 1.6 (syst.) \pm 0.9 (lumi.) \quad 
\mathrm{pb} ,
\nonumber
\end{eqnarray}
which is consistent with the SM expectations \cite{smwpair}.

Inclusive W$\gamma$ and Z$\gamma$ productions are also studied \cite{wgamma}.
The photon can be radiated from the initial state quark and final state lepton 
in addition to from the W boson with the triple gauge coupling 
which contribution is presumably enhanced in high $E_{T}$ region.
The $E_{T}$ spectra probes the anomaly to that triple gauge boson coupling.
In Figure \ref{wgammaphoet}, we show the photon $E_{T}$ distribution, and no 
excess is observed. The measured production cross sections are
\begin{eqnarray}
\sigma(p\bar{p} \rightarrow W\gamma + X) \cdot Br.(W \rightarrow \l\nu) = 
19.7 \pm 1.7 (stat.) \pm 2.0 (syst.) \pm 1.1 (lumi.) \quad \mathrm{pb} ,
\nonumber \\
\sigma(p\bar{p} \rightarrow Z/\gamma^{*}\gamma + X) \cdot 
Br.(Z \rightarrow \l^{+}\l^{-}) = 
5.3 \pm 0.6 (stat.) \pm 0.4 (syst.) \pm 0.3 (lumi.) \quad \mathrm{pb} .
\nonumber
\end{eqnarray}
Those results are compared with the NLO calculation \cite{baur} of 
19.3 $\pm$ 1.3 pb for W$\gamma$ production and 5.4 $\pm$ 0.4 pb for Z$\gamma$ 
production, and both results are a good agreement with the theoretical 
calculation. A more direct test of the gauge couplings can be performed if 
photon angular distributions of those events are studied and radiation 
amplitude zero is looked for directly.

\section{Top Quark Physics}

\subsection{Overview}

Since the discovery of the top quark in 1995 by the CDF and D0 
collaborations \cite{topdiscovery}, the studies of the top quark properties 
have been the most important mission in Run II physics program. While the 
existence of the top quark demonstrates correctness of the Standard Model, 
nothing tells us about the hierarchy problem of the mass gap among the quark 
family and generations. Detailed studies of their properties will reveal 
something about a fundamental nature of the matter.

At Tevatron, the production mechanism is a top and anti-top quarks pair 
creation via quark-antiquark annihilation (85\%) or via gluon fusion (15\%). 
Then, the top quark is assumed to decay into W boson and bottom quark 
immediately. The final observed signature thus depends on the W bosons decay 
from top quarks. The analysis strategy is basically categorized by the 
{\it dilepton} ($\l^{+}\nu b\l^{-}\bar{\nu}\bar{b}$, $\sim$7\% of all 
$t\bar{t}$ events), {\it lepton+jets} ($\l\nu bq\bar{q'}\bar{b}$, $\sim$35\%), 
and {\it all hadronic} ($q\bar{q'}bq\bar{q'}\bar{b}$, $\sim$44\%) channels. 
The analysis is also categorized by the b-tagging algorithms, but combined 
results are only presented in this section.

\subsection{Production cross sections}

\begin{figure}[hptb]
\begin{multicols}{2}
\includegraphics[width=7.0cm]{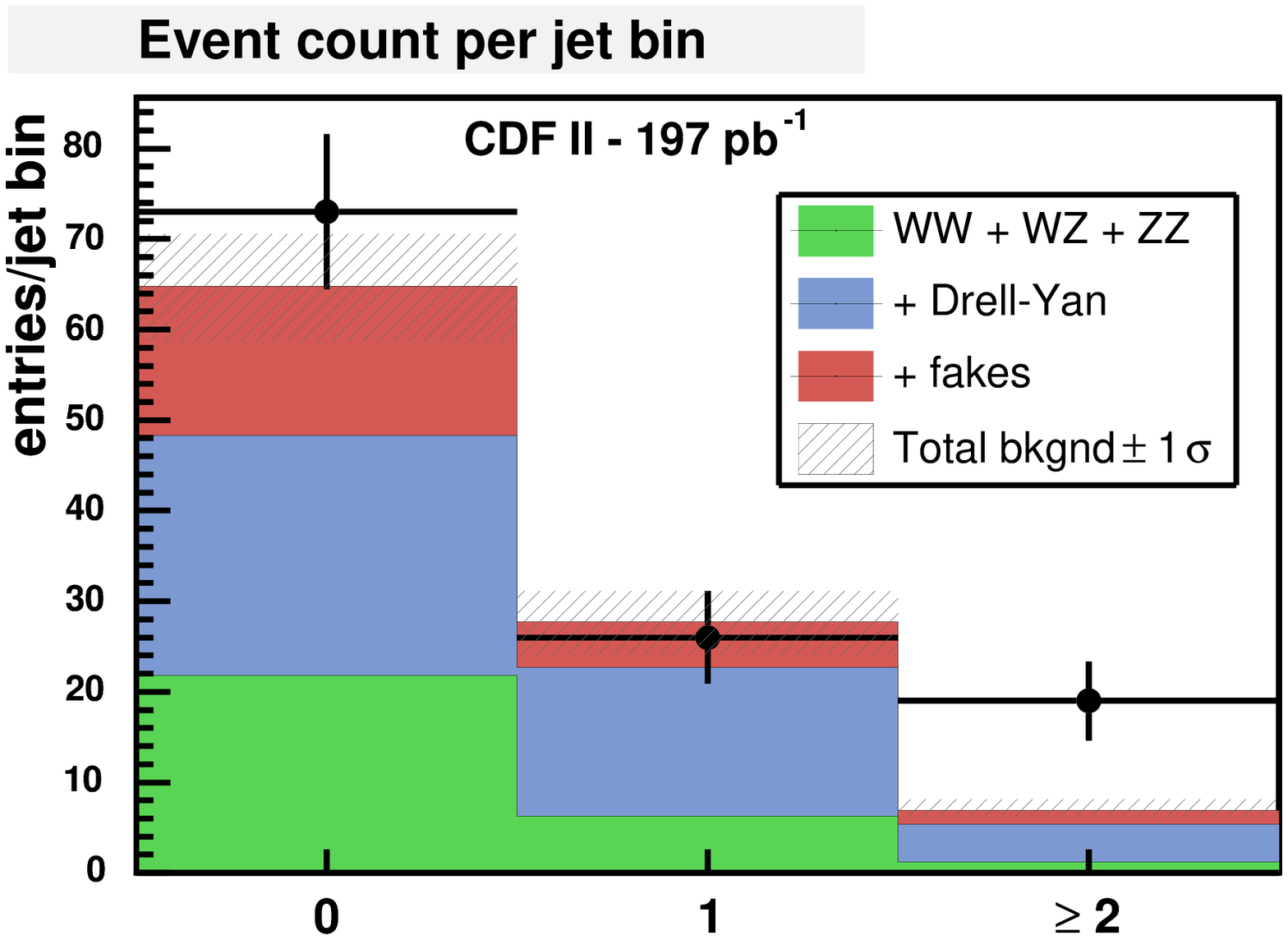}
\caption{Jet multiplicity distribution of top candidate events in the 
{\it dilepton} channel (Lepton+track).}
\label{topdilepton}
\includegraphics[width=7.0cm]{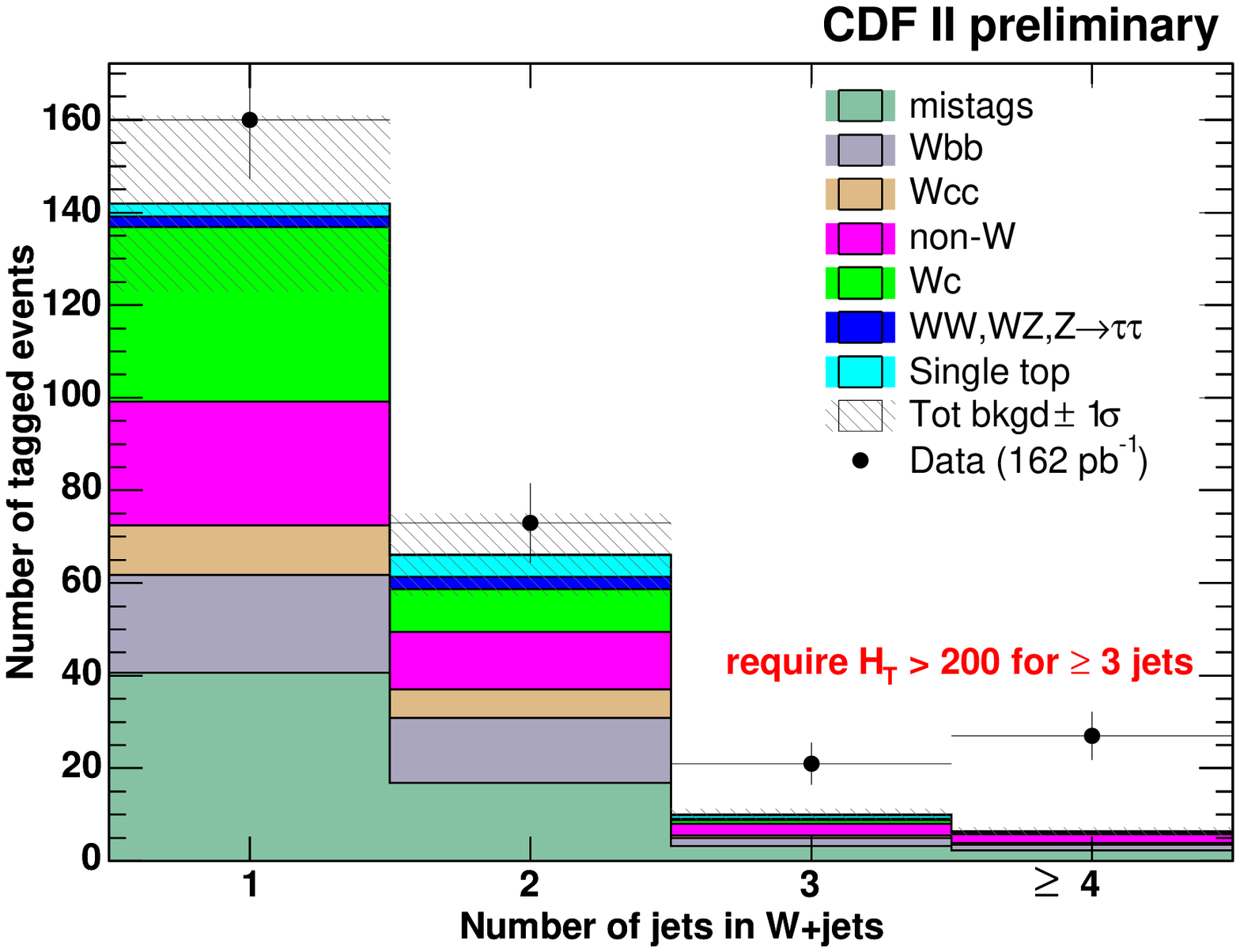}
\caption{Jet multiplicity distribution of top candidate events in the 
{\it lepton+jet} channel with Secondary Vertex b-tagging.}
\label{secvtxttprod}
\end{multicols}
\end{figure}

The {\it dilepton} channel is characterized by two leptons with high 
transverse momentum and missing energy from the undetected neutrinos and two 
jets from the b quarks. This channel is relatively clean to the other 
channels, but statistically limited. To increase the overall acceptance of two 
leptons, the selection criteria to identify the leptons is loosen, while the 
tight selection does not suffer from the background contamination. 
Figure \ref{topdilepton} shows the jet multiplicity distribution of the 
candidate events in case of one tight lepton plus isolated track data set.
Those combined result for the production cross section is \cite{ttdilep}
\begin{eqnarray}
\sigma(p\bar{p} \rightarrow t\bar{t} + X) = 
7.0 ^{+2.4}_{-2.1} (stat.) ^{+1.6}_{-1.1} (syst.) 
\pm 0.4 (lumi.) \quad \mathrm{pb} ,
\nonumber
\end{eqnarray}
in good agreement with a theory prediction of 6.7$^{+0.7}_{-0.9}$ pb at NLO 
level with assuming the top mass of 175 GeV \cite{topnlo}.

The {\it lepton+jet} channel requires one lepton and 4 more jets in the final 
state. The sample size is larger than the {\it dilepton} sample, but there is 
a significant background contamination from the associated production of W 
boson with jets. The purity of the sample can be improved by the 
identification of at least one b-quark jets. We typically use two different 
b-tagging algorithms. The first one makes used of the long lifetime of the b 
hadrons to reconstruct a displaced secondary vertex. The second one identifies 
the soft muon from semileptonic b-decay. The signal to noise ratio can be 
further improved by requiring the scalar sum of the energy in the event, 
$H_{T}$, to be greater than 200 GeV. The jet multiplicity distribution of 
these candidate events by the displaced secondary vertex technique with 
$H_{T}$ $>$ 200 GeV are shown in Figure \ref{secvtxttprod}. The excess of 
events in the bins $N_{jet}$ $\geq$ 3 bins is nicely described after the 
inclusion of top contributions. The production cross sections from both 
b-tagging algorithms are measured to be \cite{xsecvtx,xsecslt}
\begin{eqnarray}
\sigma(p\bar{p} \rightarrow t\bar{t} + X) & = 
5.6 ^{+1.2}_{-1.1} (stat.) ^{+0.9}_{-0.6} (syst.) \quad \mathrm{pb} 
\quad \mathrm{(Secondary \; Vertex)}, \nonumber \\
 & = 5.2 ^{+2.9}_{-1.9} (stat.) ^{+1.3}_{-1.0} (syst.) \quad \mathrm{pb} 
\quad \mathrm{(Soft \; Lepton)}. \qquad \; \nonumber
\end{eqnarray}

\begin{figure}[hptb]
\begin{multicols}{2}
\includegraphics[width=8.0cm]{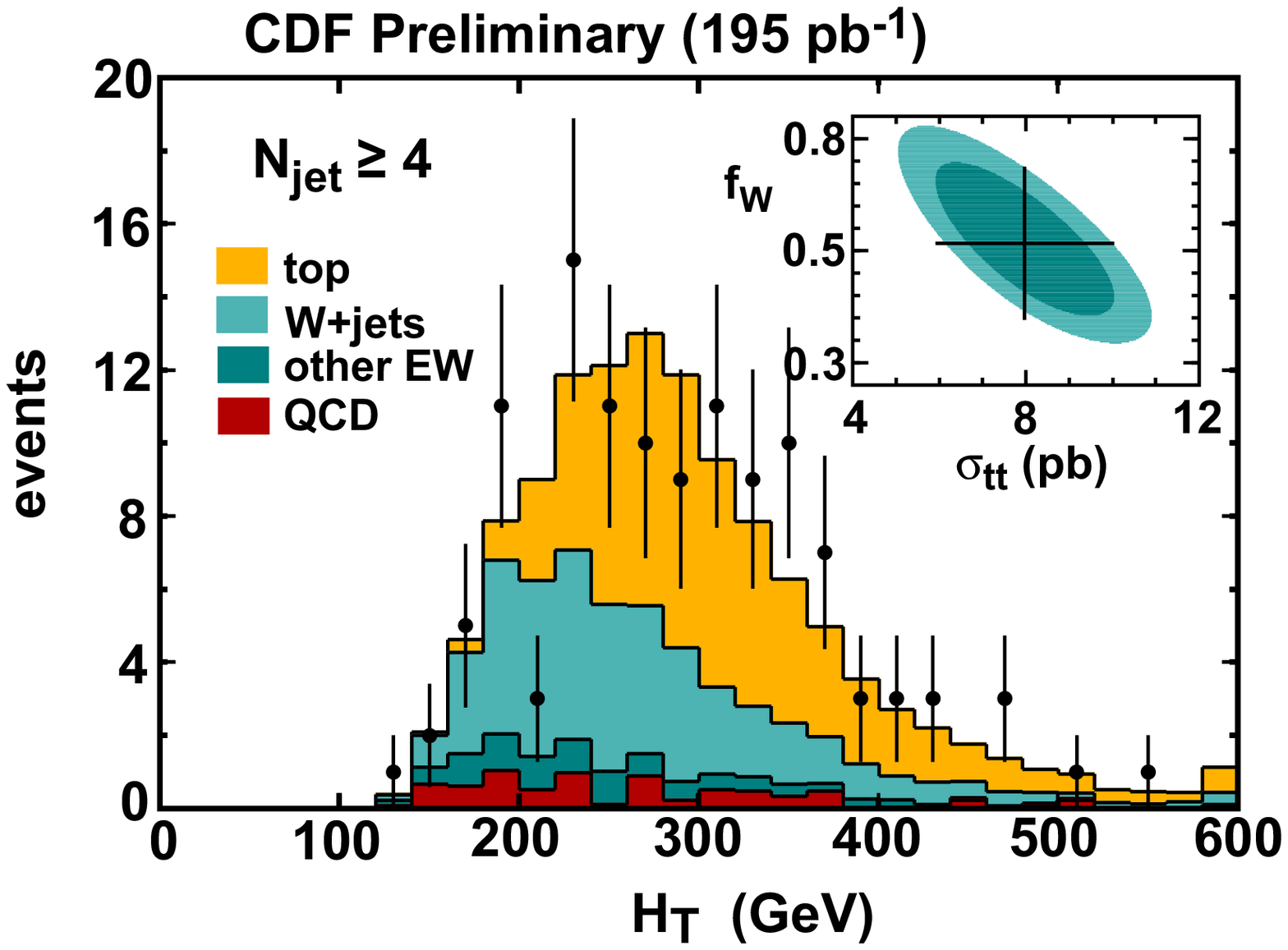}
\caption{$H_{T}$ distribution with the expected shapes of signal and 
backgrounds.}
\label{htdis}
\hspace*{1cm}
\includegraphics[width=5.5cm]{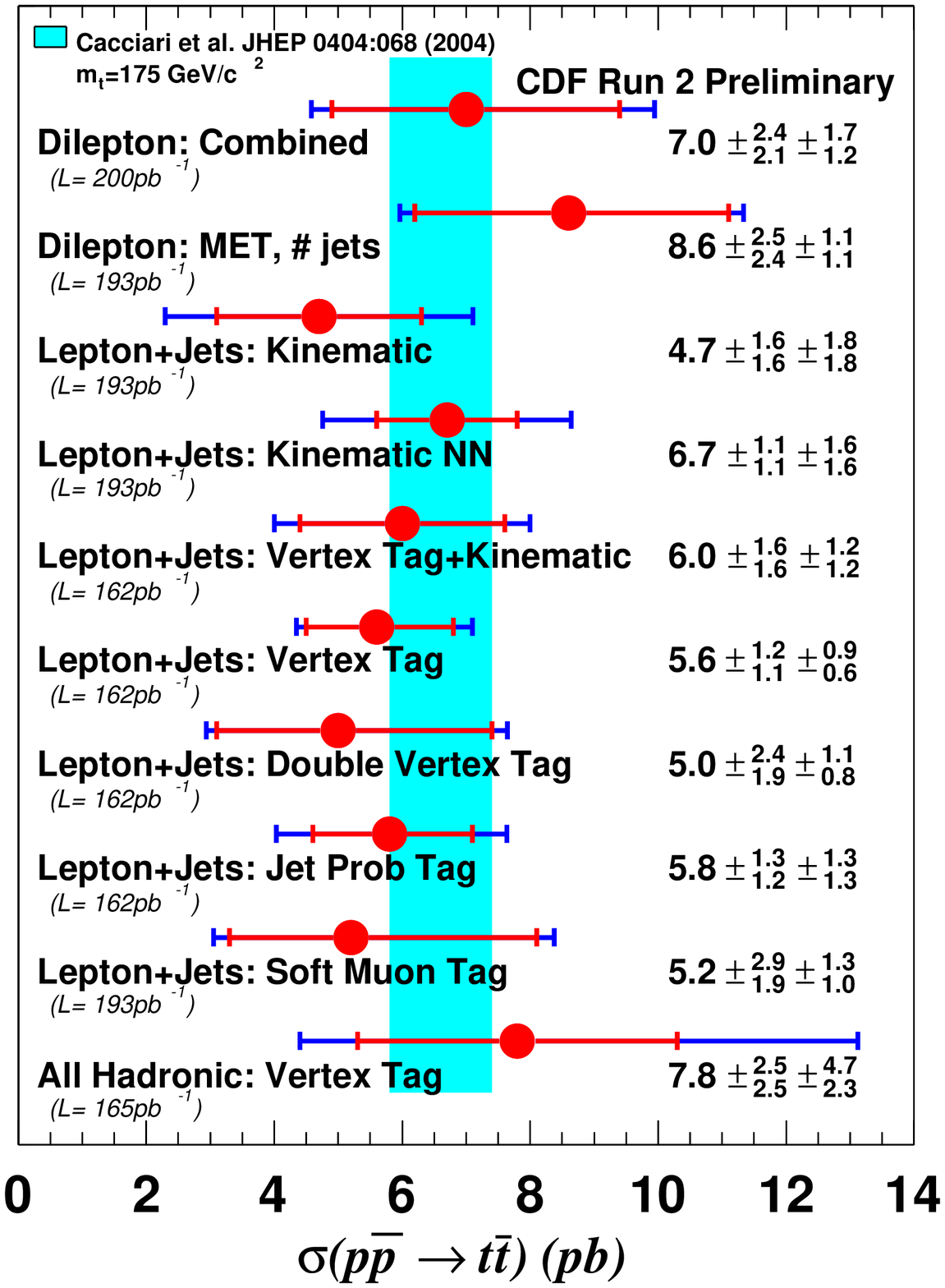}
\caption{Summary of the measured $t\bar{t}$ cross section from CDF.}
\label{allxsec}
\end{multicols}
\end{figure}

Instead of counting signal and background events, one can extract the fraction 
of $t\bar{t}$ events in the lepton + jets sample by fitting one or more 
kinematic variables in the data to the expected shapes from signal and 
backgrounds. As already mentioned, the $H_{T}$ variable is strong 
discrimination power of signal to backgrounds. Figure \ref{htdis} shows the 
$H_{T}$ distribution with the expected shapes of signal and backgrounds. At 
least 4 more jets are required for the fitting to $H_{T}$ in this plot. The 
estimated production cross section is \cite{htfit}
\begin{eqnarray}
\sigma(p\bar{p} \rightarrow t\bar{t} + X) = 
4.7 \pm 1.6 (stat.) \pm 1.8 (syst.) \quad \mathrm{pb} .
\nonumber
\end{eqnarray}

Summary of the measured $t\bar{t}$ cross section from CDF is presented in 
Figure \ref{allxsec}. Note that there are many other measurements using 
various different methods \cite{otherxsec}. 

\subsection{Top quark mass}

Recent combined result of the top quark mass measurements by CDF and D0 
collaboration in Run I experiment is reported to be 178.0 $\pm$ 
4.3 GeV \cite{0404010}. Global EW fits place a 95\% CL upper bound on the 
Higgs mass of $\sim$260 GeV \cite{lepew}. The precise measurement of the top 
quark mass will further constrain the mass bound. The Run II goal of the mass 
measurement is to control under an uncertainty of 2-3 GeV.

The top mass measurement at CDF largely depends on the fitting techniques to
reconstruct the event topology. First, the kinematic equations of the 
$t\bar{t}$ decay chain are imposed, and then some kinematical variables are 
fold by the measured quantities out of 16 free parameters in each event. 
In {\it lepton+jets} channel, there are 24 configurations for the jets.
For each configuration, an event $\chi^{2}$ is calculated and minimized. The 
$\chi^{2}$ takes into account the detector resolution on the measured 
quantities as well as the W boson and top quark widths. The configuration that 
yields the minimum $\chi^{2}$ is taken as the configuration for the event. The 
distribution of the reconstructed mass is compared and fit to Monte Carlo 
generated templates with known input top masses.

\begin{figure}[hptb]
\begin{multicols}{2}
\includegraphics[width=8.0cm]{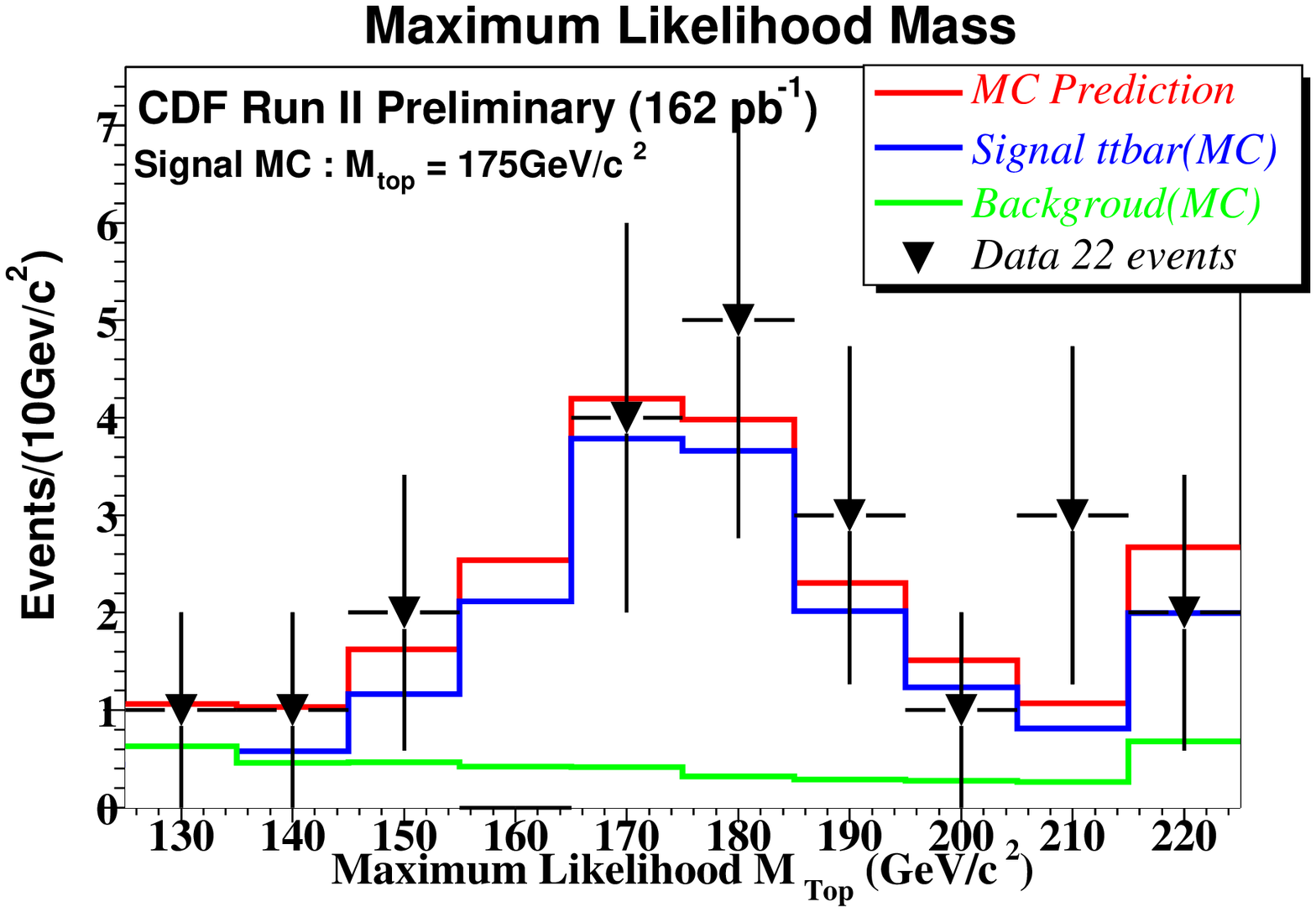}
\caption{Maximum likelihood mass distribution by DLM analysis.}
\label{dlmmass}
\hspace*{1cm}
\includegraphics[width=5.0cm]{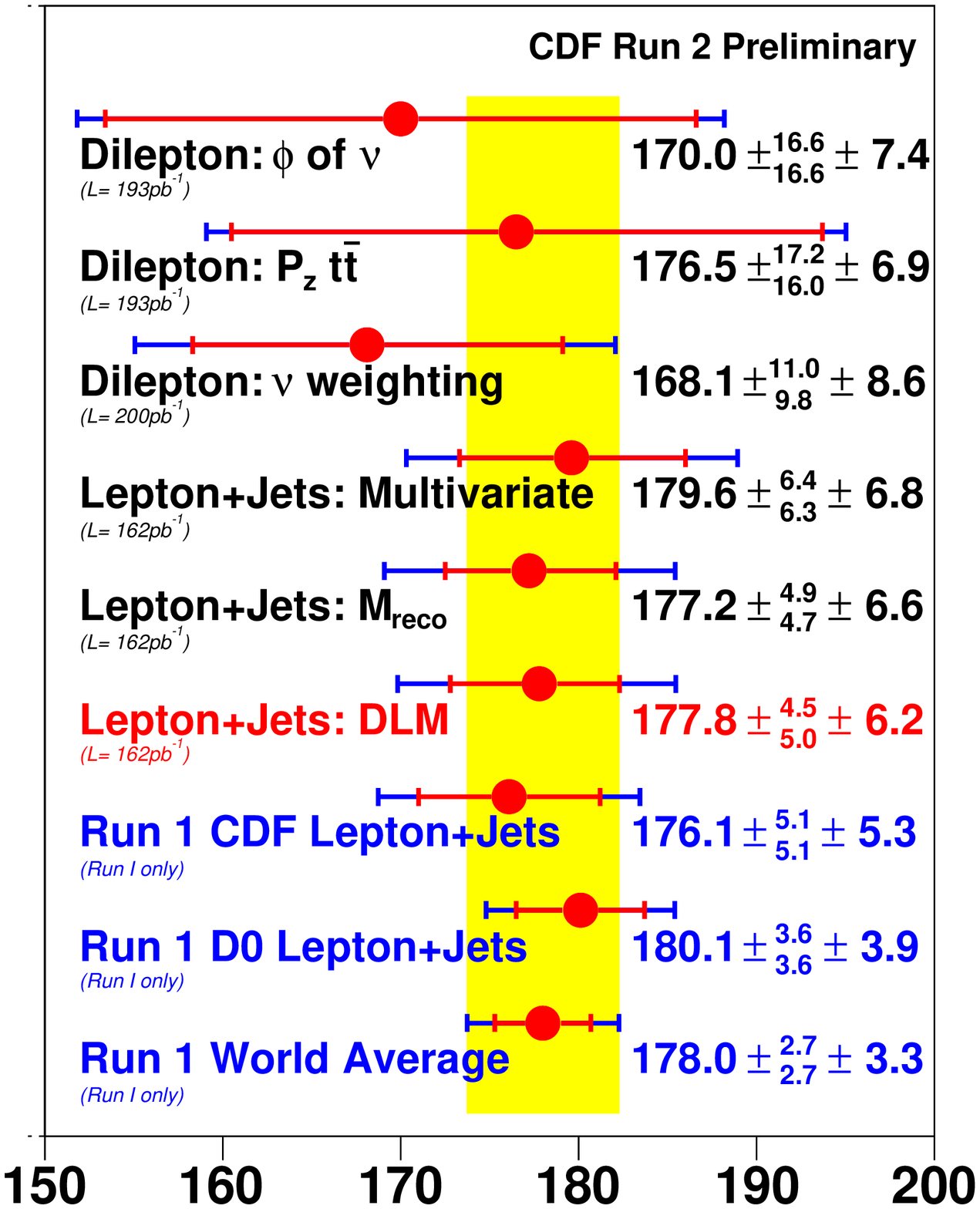}
\caption{Summary of the measured top mass from CDF.}
\label{alltmass}
\end{multicols}
\end{figure}

The CDF has measured the mass of the top quark in {\it lepton+jets} channel 
using traditional ``template'' methods \cite{ksato}, where one mass is 
reconstructed per event and the resulting mass distribution compared against 
template distributions from the simulated $t\bar{t}$ events of varying masses. 
The multivariate templates method \cite{mtm} is used weighting events 
according to the probability for the chosen jet-parton assignment to be 
correct. The Dynamical Likelihood Method \cite{kondo} uses the probability 
formed from the $t\bar{t}$ matrix element to utilize maximal information from 
the events. This method \cite{yorita} results in the best value of the top 
quark mass of
\begin{eqnarray}
M_{t} = 177.8 ^{+4.5}_{-5.0} (stat.) \pm 6.2 (syst.) \quad \mathrm{GeV} .
\nonumber
\end{eqnarray}
Figure \ref{dlmmass} shows the top mass distribution by DLM analysis. The 
systematic uncertainty is dominated by the measurement of the jet energy.

In {\it dilepton} channel, due to the presence of two unfold neutrinos, one 
more extra conditions driven by the Monte Carlo simulation are 
imposed on the kinematical variables for the neutrinos. The best fit returns 
\begin{eqnarray}
M_{t} = 168.1 ^{+11.0}_{-9.8} (stat.) \pm 8.6 (syst.) \quad \mathrm{GeV} ,
\nonumber
\end{eqnarray}
by the neutrino weighting method \cite{nwm}. The other top mass measurements 
using {\it dilepton} channel can be seen in \cite{dilepms}.

Summary of the top mass measurement from CDF is presented in 
Figure \ref{alltmass}. 

\subsection{W helicity in top decay}

The W boson coming from top decay can be either left-handed (L) or 
longitudinally (0) polarized. Those fraction for both mixture states is 
predicted in the Standard Model as
\begin{eqnarray}
f_{0} = \frac{\Gamma(t \rightarrow bW_{0})}
{\Gamma(t \rightarrow bW_{0}) + \Gamma(t \rightarrow bW_{L})} = 
\frac{M_{t}^{2}}{M_{t}^{2} + 2 M_{W}^{2}} = 0.70 ,
\nonumber
\end{eqnarray}
where b-quark mass is neglected and top mass is assumed to be 175 GeV. It is 
important to check this fraction of the longitudinal component which tells us 
the symmetry breaking mechanism to give a mass of W boson.

Since the polarization state of the W boson controls the angular distribution 
of the decaying leptons, CDF has precisely measured the lepton $p_{T}$ spectra 
and angular distributions. The estimated $f_{0}$ value from those 
distributions constrains to be \cite{whel1,whel2}
\begin{eqnarray}
0.25 < f_{0} < 0.88 \qquad \mathrm{at} \; 95\% \; \mathrm{C.L.} ,
\nonumber
\end{eqnarray}
which is consistent range of the Standard Model prediction.

\subsection{Search for single top production}

The single top production is linked to the direct measurement of the quark 
mixing among third generation, $|V_{tb}|^{2}$, in CKM matrix \cite{ckm}.
The dominant production mechanics are s-channel Drell-Yan production and 
t-channel associated W boson scattering process. Those production cross 
sections are proportional to the $|V_{tb}|^{2}$. If abnormal cross section is 
observed, it is directory addressed to the coupling with the Wtb electroweak 
interaction. 

However, the single top event has lower jet multiplicities than that of 
$t\bar{t}$ production events, which suffers significant backgrounds from W + 
jets events. The optimized search is proposed in \cite{singletopopt}. The CDF 
has searched the single top candidate events, and no candidate observed. We 
set a limit of \cite{singletoprunii}
\begin{eqnarray}
\sigma(p\bar{p} \rightarrow tb + X) & < 17.8 
\qquad \mathrm{at} \; 95\% \; \mathrm{C.L.} .
\nonumber
\end{eqnarray}

\section{Beyond the Standard Model}

\subsection{Overview}

There are several reasons to motivate a new physics beyond the Standard Model: 
the electroweak symmetry breaking, hierarchy problems, gravitational force, 
and so forth. Those issues inspire a new theory beyond the Standard Model such 
as Supersymmetry (SUSY), Grand Unified Theories (GUT), Technicolor (TC), and 
Extra Dimensions (ED). 

These models predict new signatures that can be seen at the experiments with 
a small production cross section to be compared with the typical QCD events at 
Tevatron. It is often hard to distinguish new phenomena from the SM 
background processes. Jet based strategies are overwhelmed by the SM 
processes, so that the CDF has searched lepton based signatures even through 
the rates are often suppressed. In this section, a couple of selected studies 
are only described. Readers may refer \cite{exsotics} for details.

\subsection{Search for narrow resonance in high mass region}

\begin{figure}[hptb]
\begin{multicols}{2}
\includegraphics[width=6.0cm]{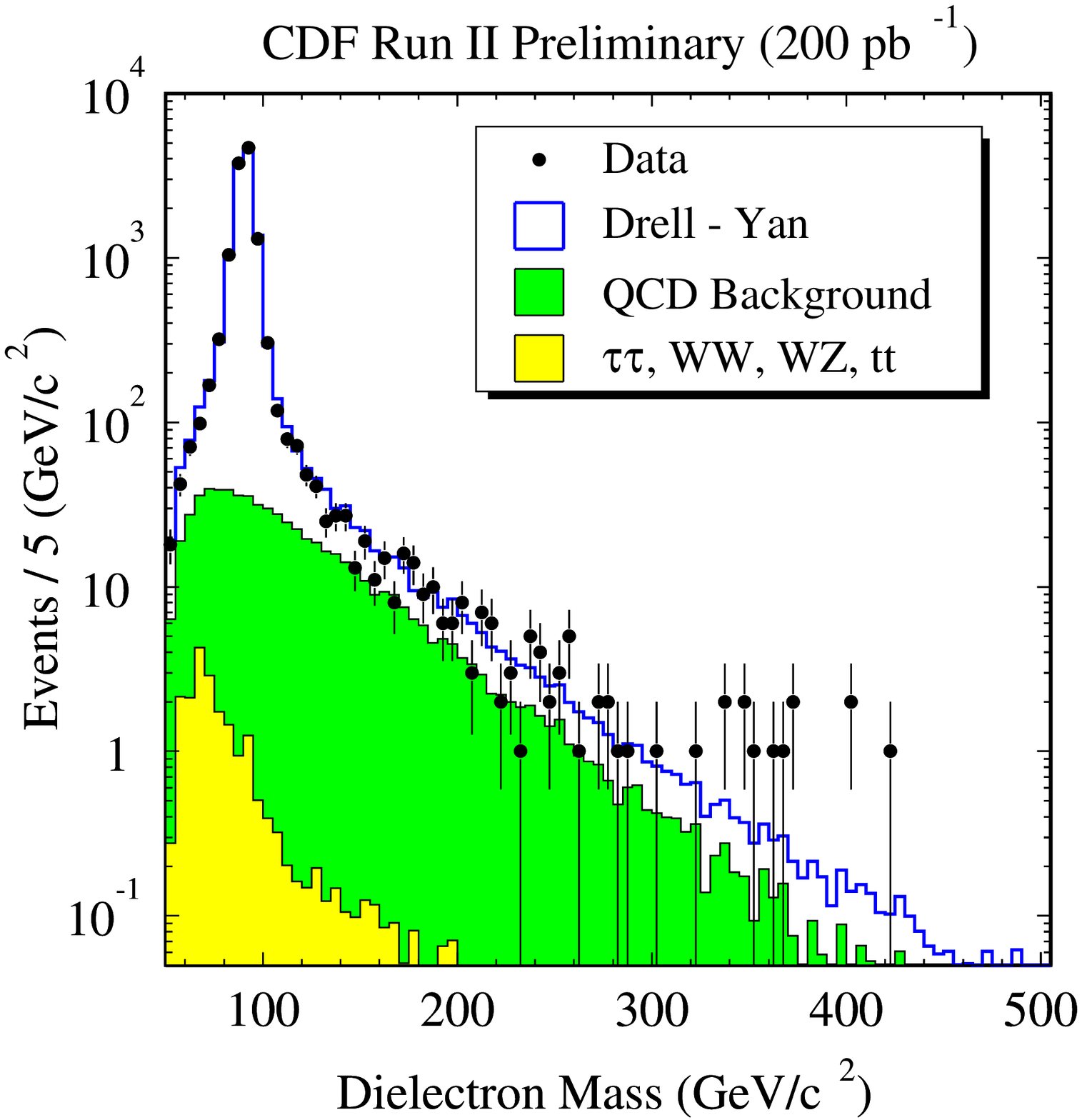}
\caption{Invariant mass distribution of {\it dilepton} events.}
\label{dyeemass}
\includegraphics[width=6.0cm]{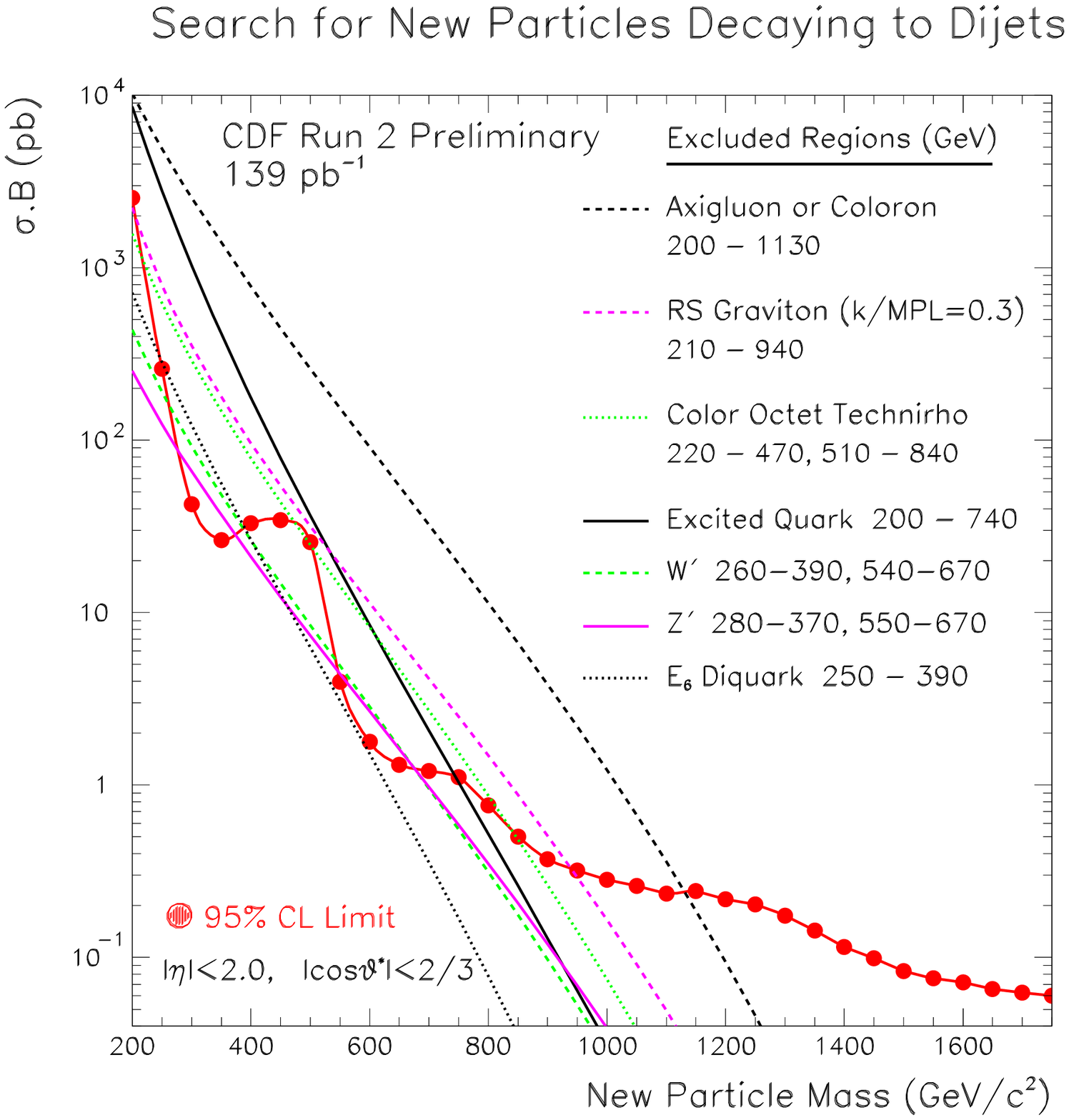}
\caption{The 95\% C.L. upper limit on the cross section times branching ratio 
for new particles decaying {\it dijet} for various models.}
\label{dijetmass}
\end{multicols}
\end{figure}

A heavy partner of the Z boson, Z' boson \cite{zprime}, is a by-product 
predicted by many models of GUT, ED, and little Higgs \cite{littlehiggs} 
models. The CDF has searched the excess in high mass Drell-Yan process
using {\it dilepton} events. The invariant mass distribution of {\it dilepton} 
events is shown in Figure \ref{dyeemass}. No excess is found, and limits is 
set the cross section limits as a function of the spin state since the 
experimental acceptance changes depending their spin state of the resonance 
particles. Mass limits for the RPV $\tilde{\nu}$ (spin-0), 
Z' \cite{so10,e6,littlehiggs,tc} (spin-1), and RS-graviton \cite{rs} (spin-2) 
are set \cite{ikado}.

The {\it dijet} events are also used to search for the high mass resonance 
peak, and limits for some models are set from the cross section limit as shown 
in Figure \ref{dijetmass}.

\subsection{Search for Higgs boson(s)}

With high statistic data in Run II, CDF and D0 experiments will start reaching 
sensitivity to production of low-mass Higgs bosons beyond the limits from LEP 
experiment. The Standard Model Higgs boson is predominantly decaying into 
$b\bar{b}$ in the low-mass region. The most promising channel is the 
associated production with W or Z boson followed by leptonically decay. In 
high-mass region up to about 180 GeV, Higgs boson produced via gluon fusion 
might be observable in the decay to WW. In Higgs doublet model, the coupling 
of the Higgs bosons to b-quarks can be significantly enhanced which allows us 
to search for the Higgs bosons produced in association with b-quarks. The 
previous sensitivity studies can be seen in \cite{higgs1,higgs2}.

\begin{figure}[hptb]
\begin{multicols}{2}
\includegraphics[width=7.0cm]{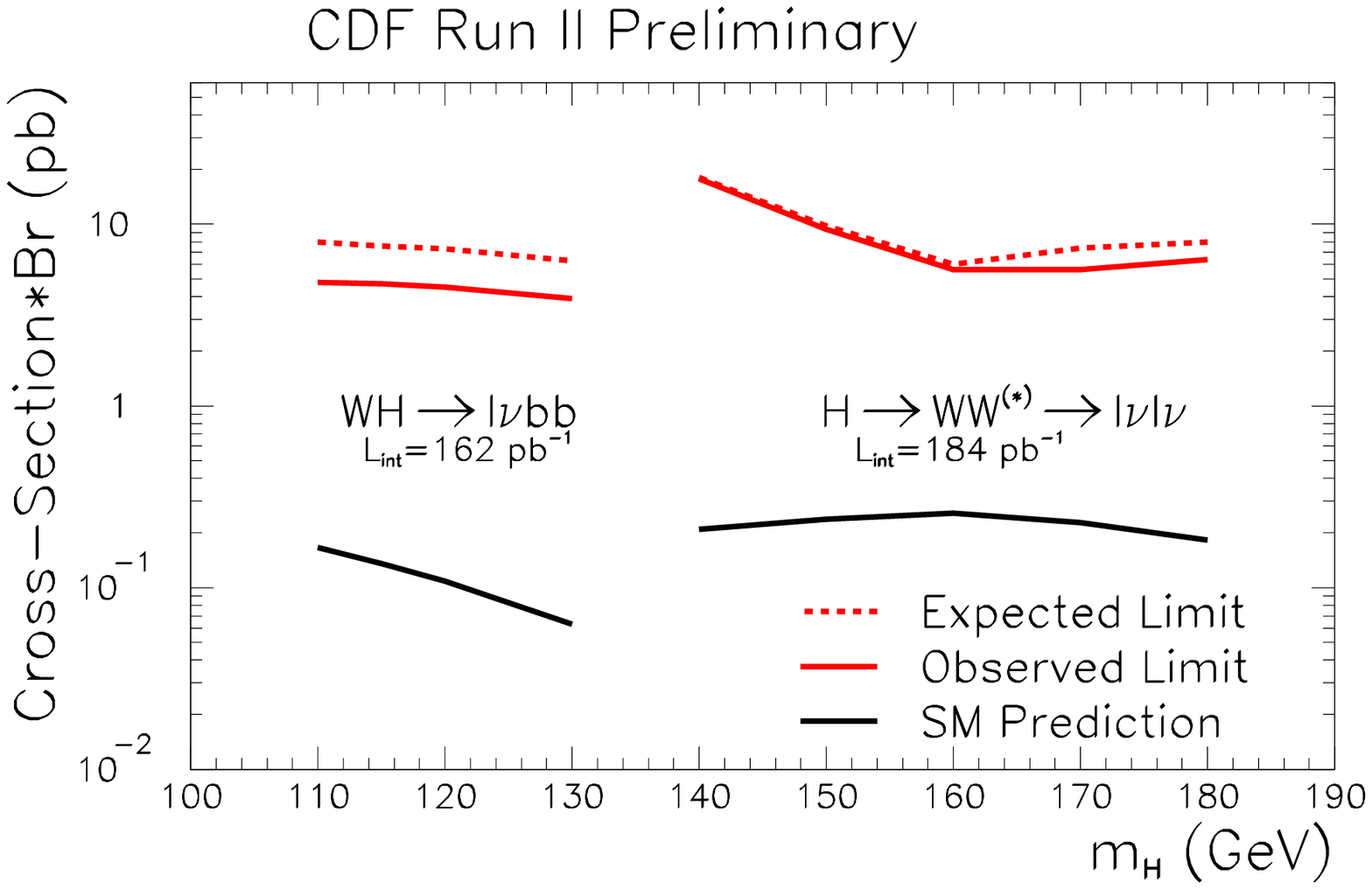}
\caption{Cross section times branching ratio limit as a function of SM Higgs 
boson mass.}
\label{hsen}
\includegraphics[width=5.0cm]{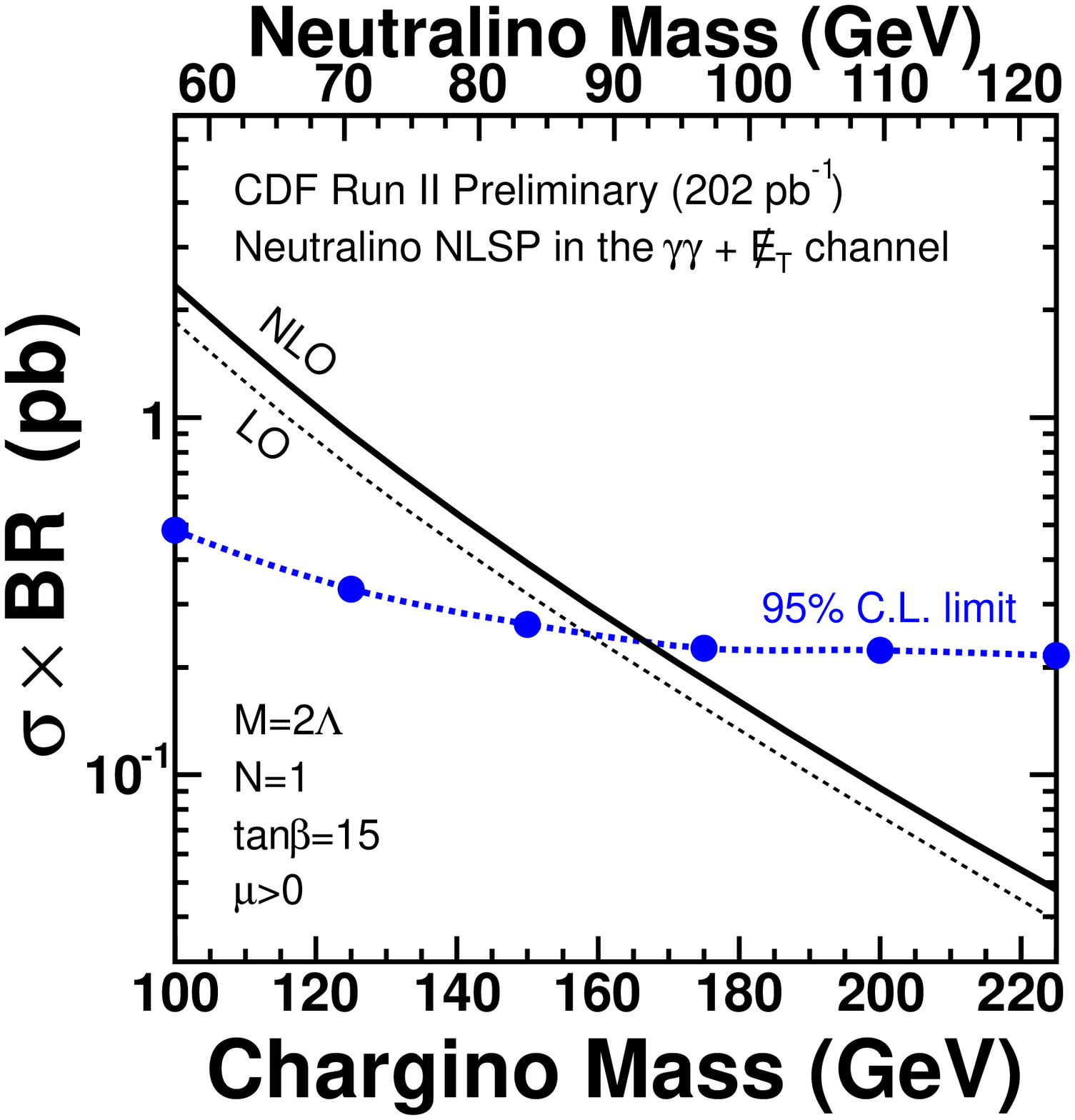}
\caption{Cross section times branching ratio limit as a function of chargino 
mass.}
\label{gmsbsen}
\end{multicols}
\end{figure}

The CDF has searched the Higgs boson(s), and set the cross section limits 
as a function of Higgs mass for the searches of the associated production with 
W boson ($\l\nu b\bar{b}$) \cite{ishizawa} and Higgs WW decay channel 
($\l\nu\l\nu$ or $\l\nu\l\nu\l\nu$) \cite{kobayashi}. The analysis strategies 
are the same as the top quark measurement in {\it lepton + jets} channel and 
the WW cross section measurement. The sensitivity plot is shown in 
Figure \ref{hsen}. Also, Many other Higgs searches are on-going.

\subsection{Search for SUSY in {\it diphoton + missing $E_{T}$} channel}

Introduction of supersymmetric particles has attractive features for lots of 
aspects: Higgs self-coupling is converged without fine tuning, all forces 
except gravity merging in GUT scale, and resulting in many new particles in 
TeV region, even in cosmology. In this section, a search of Gauge-Mediated 
SUSY breaking (GMSB) scenario is only described. Many results can be seen 
in \cite{exsotics}.

GMSB can produce a final state of two photons and large missing energy 
from the lightest supersymmetric particle (LSP), where the lightest neutralino 
can decay into photon and gravitino (LSP). Thus, {\it diphoton} candidate 
events is used in this analysis. The dominant background source is QCD jets 
events which makes fake missing $E_{T}$ and photon signals due to the 
mismeasurement of jets and interaction vertices. After requiring large missing 
$E_{T}$ with two identified photons, no candidate is found, and thus set 
limits \cite{gmsbana}, 
\begin{eqnarray}
m_{\tilde{\chi}^{\pm}_{1}} > 93 \; \mathrm{GeV} \quad \mathrm{and} \quad 
\Lambda > 69 \; \mathrm{TeV} \qquad \mathrm{at} \; 95\% \; \mathrm{C.L.} ,
\nonumber
\end{eqnarray}
where $\Lambda$ is a SUSY-breaking scale. The sensitivity plot is shown in 
Figure \ref{gmsbsen}.

\subsection{Search for Leptoquarks}

The Leptoquark (LQ) model \cite{lqmodel} is natural consequence from GUT 
theories. The relation between quark and lepton quantum numbers can rule out a 
triangle anomaly \cite{triangleanomarly}, so that the theory is 
renormalizable. The Leptoquarks are color triplet bosons carrying both lepton 
and quark quantum numbers, they can be scalar or vector bosons. At Tevatron, 
they can be pair produced through strong interactions and decay either into a 
charge lepton and a quark ($\beta$ = 1) or a neutrino and a quark 
($\beta$ = 0). 

\begin{figure}[hptb]
\begin{multicols}{2}
\includegraphics[width=7.0cm]{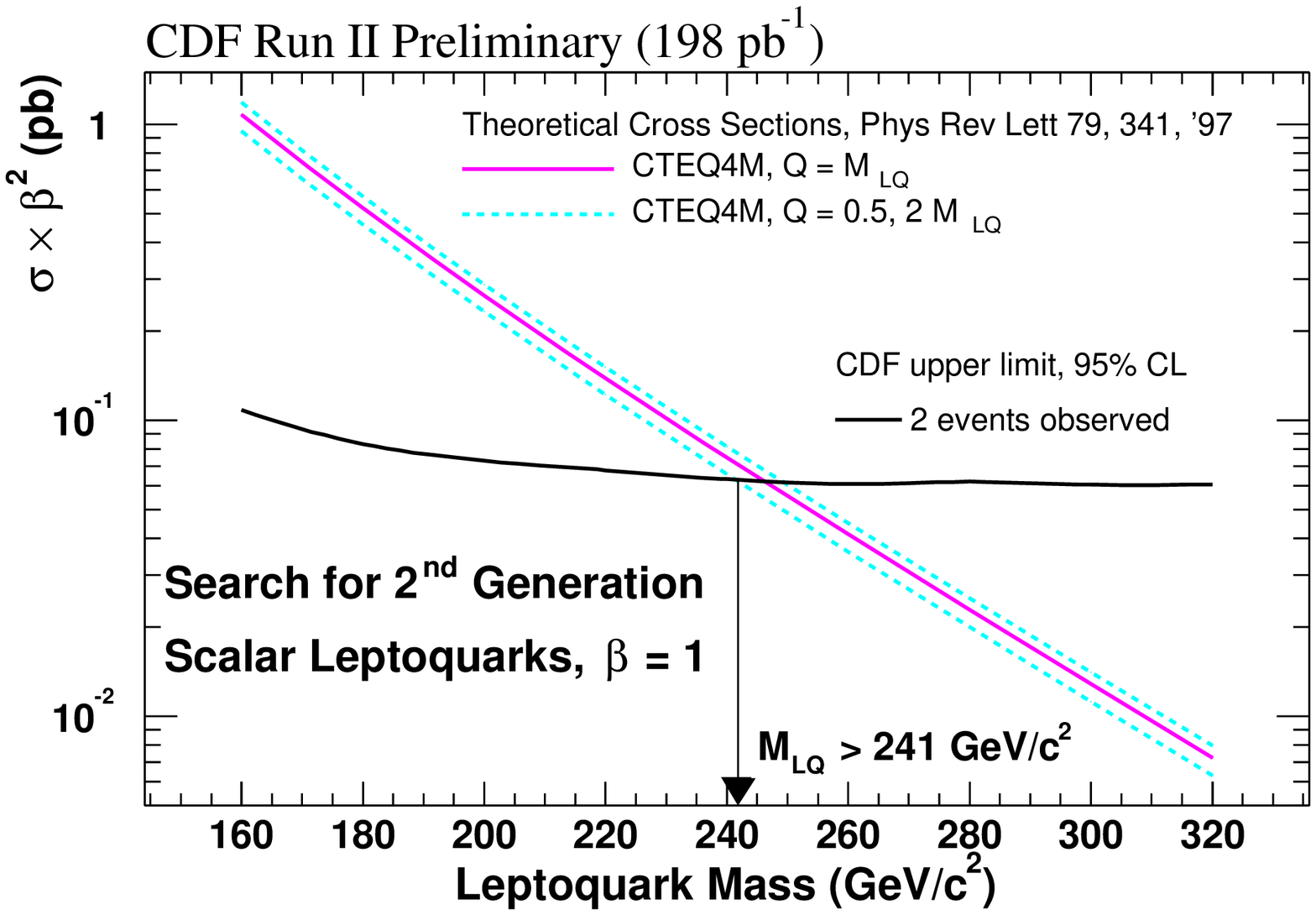}
\caption{Upper limit on the cross section times squared branching ratio for 
scalar leptoquark production ($\beta$ = 1).}
\label{leptoquarkbeta1}
\includegraphics[width=6.0cm]{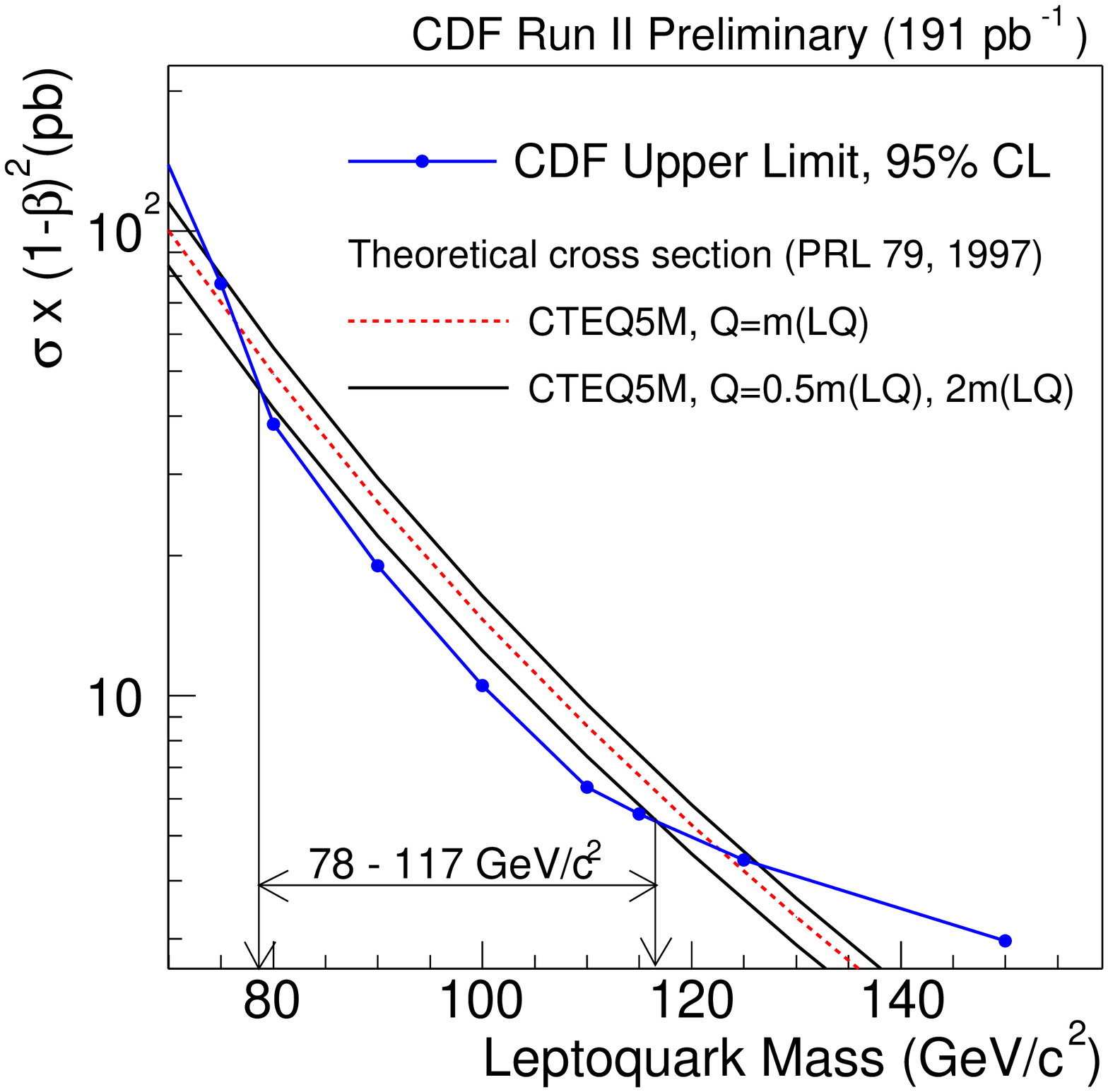}
\caption{Upper limit on the cross section times squared branching ratio for 
scalar leptoquark production ($\beta$ = 0).}
\label{leptoquarkbeta2}
\end{multicols}
\end{figure}

The CDF has searched two scenarios of $\beta$ = 1 and $\beta$ = 0 for the 
scalar type Leptoquarks. Final signatures are $\l^{+}\l^{-}q\bar{q}$ 
($\beta$ = 1), $\nu\bar{\nu}q\bar{q}$ ($\beta$ = 0), and $\l\nu q\bar{q}$ 
($\beta$ = 0.5) for the first and second generations of the scalar 
Leptoquarks. The results are shown in Figure \ref{leptoquarkbeta1} and 
\ref{leptoquarkbeta2} in case of $\beta$ = 1 and $\beta$ = 0, respectively, 
and limits is set \cite{leptana1}. A search for the third generation 
Leptoquark is also in progress.

\subsection{Search for Excited Electron}

\begin{figure}[hptb]
\begin{multicols}{2}
\includegraphics[width=6.0cm]{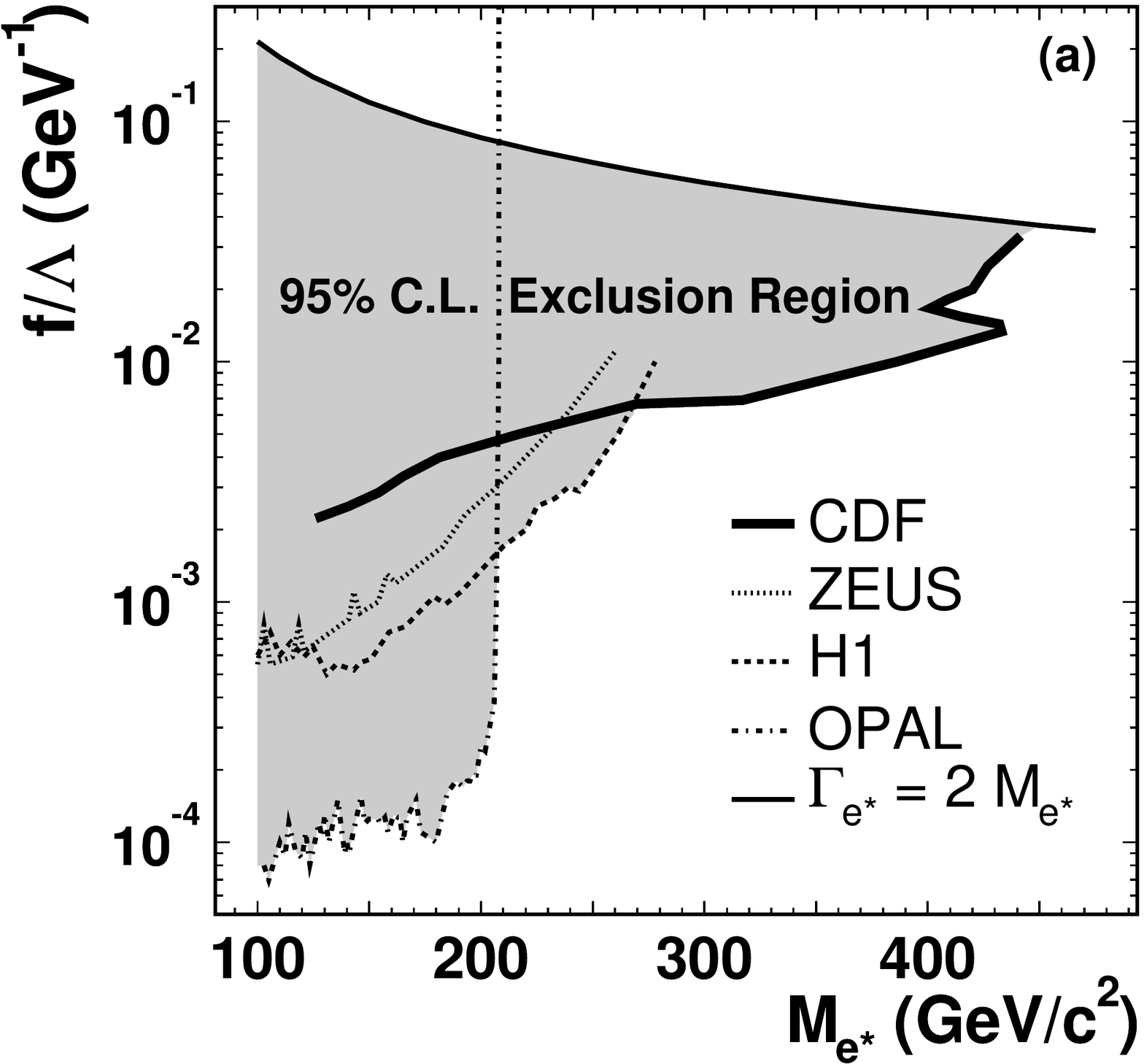}
\caption{The 95\% C.L. excluded region in the parameter space of the 
gauge-mediated model.}
\label{excitedelegm}
\includegraphics[width=6.0cm]{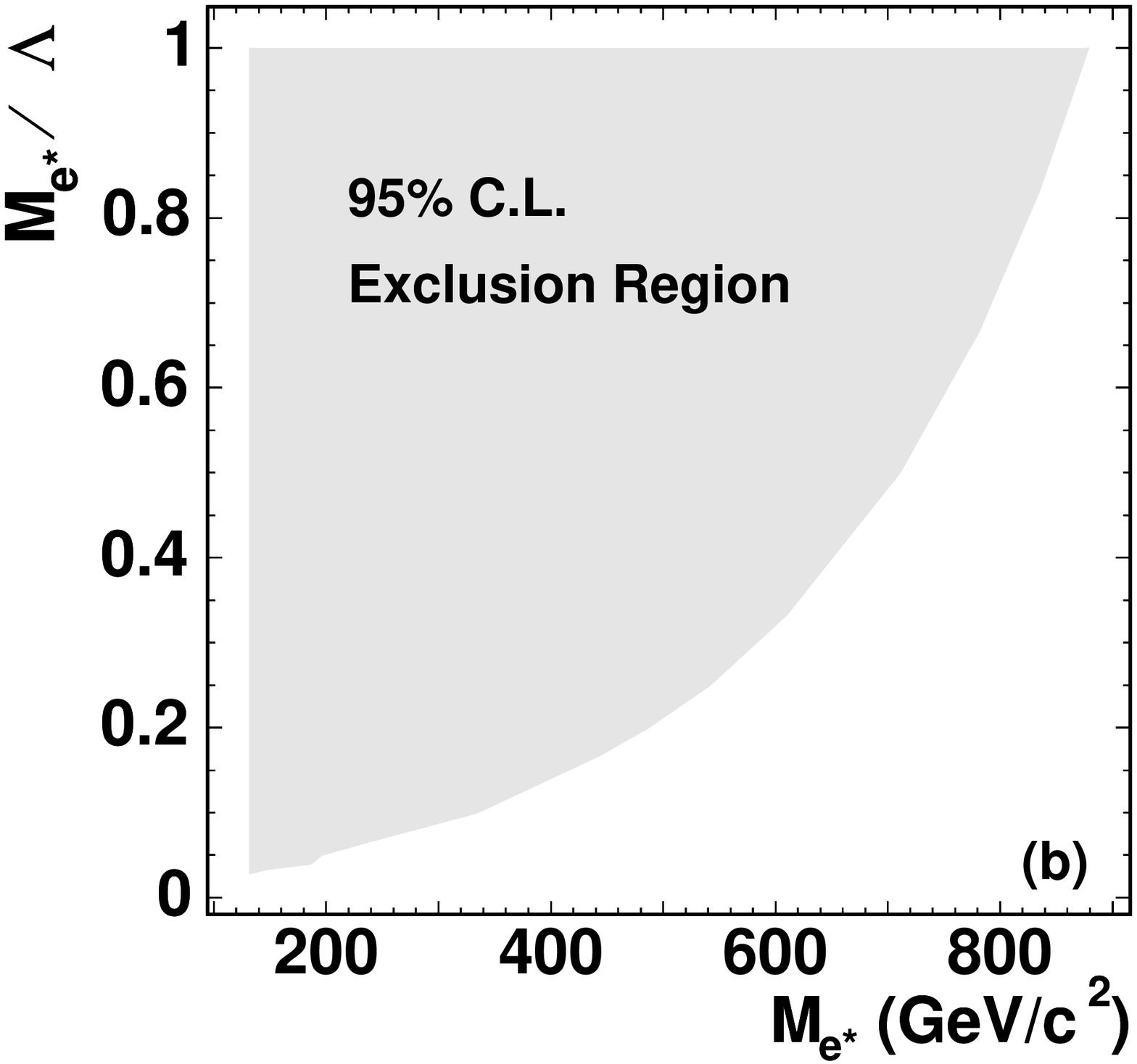}
\caption{The 95\% C.L. excluded region in the parameter space of the contact 
interaction model.}
\label{excitedeleci}
\end{multicols}
\end{figure}

In the compositeness model for the substructure of quarks and leptons, 
the excited states of those particles are considerable \cite{excitedele}. The 
CDF has searched the excited electron, $e^{*}$, which could be produced 
through either contact interaction or gauge mediated interactions, and decays 
into an electron and photon. The signature would be events with $ee\gamma$ 
with making a mass peak of the excited electron with $M_{e\gamma}$. There is 
no observation of any deviation from the Standard Model backgrounds. The upper 
limits are set at 95\% C.L. in the parameter space of the excited electron 
mass and the $M_{e*}$/$\Lambda$ for the contact interaction model or 
$f$/$\lambda$ for the gauge mediated model \cite{excitedelecdf}. 
Figure \ref{excitedelegm} and \ref{excitedeleci} shows the excluded region for 
the two models, respectively.

\section{Conclusion}

The Tevatron Run II program is in progress since 2001. The CDF experiment 
have accumulated roughly five times more data than in Run I, with much 
improved detectors. Many physics results have been produced and more are 
expected in the near future. Preliminary results on the measurements at CDF 
are presented, most of all will be published in the near future.

For the electroweak physics, production of weak vector bosons has been 
measured at a new center of mass energy. The single boson productions have 
studied, and some fundamental parameters are very precisely driven. The W 
boson mass will be measured with the accuracy of 25 MeV with a 
well-established knowledge of the detector. Pairs of gauge bosons are now 
being produced in reasonably high statistics, and interactions among gauge 
bosons will be studied in detail. 

For the top physics, the top quark measurements have been re-established since 
Run I experiment, and new measurements just came out in Run II experiment. The 
top cross section and mass measurements have improved systematics with various 
analysis techniques. With large data samples, the top mass will be measured 
with the accuracy of 2 GeV. The studies of the top quark property will push 
forward with the understandings of top physics.

For new physics searches, many results are improved and marked new limits. In 
the next a few years, the Tevatron is an unique opportunity to directly search 
for new physics beyond the Standard Model. The CDF is aggressively looking for 
new phenomena.

Many other studies not fully covered in this paper are on-going. We hope to 
see many exciting results from the Tevatron.

\section{Acknowledgments}

The author would like to thank all the people in the CDF Collaboration, and 
especially Dr. Pasha Murat for the assistance that he provided the information 
during the Conference. The author would also like to thank the organizers of 
the Conference.




\end{document}